%% file: main.tex
\documentclass[a4paper]{article}

\usepackage[utf8]{inputenc}             % para los acentos
\usepackage{amsmath,amssymb,amsfonts}
\usepackage{algorithm}
\usepackage{algorithmic}
\usepackage{comment}
\usepackage{graphicx}
\usepackage{subfig}
\usepackage{textcomp}
\usepackage{xcolor}
\usepackage{listings}
\usepackage{balance}
\usepackage{soul}
\usepackage{hyperref}
\newcommand{\CommentInline}[1]{
	\hfill \textcolor{blue}{\textit{// #1 }} 
}

\newcommand\blfootnote[1]{%
  \begingroup
  \renewcommand\thefootnote{}\footnote{#1}%
  \addtocounter{footnote}{-1}%
  \endgroup
}

\usepackage{authblk}

\title{Exploring Energy Saving Opportunities in Fault Tolerant HPC Systems}
\author[1]{Marina Morán}
\author[1]{Javier Balladini}
\author[2]{Dolores Rexachs}
\author[3]{Enzo Rucci}

\affil[1]{Dpto. Ingeniería de Computadoras, Facultad de Informática, Universidad Nacional del Comahue, Argentina\\
\authorcr
\{marina.moran,javier.balladini\}@fi.uncoma.edu.ar}

\affil[2]{Dpto. Arquitectura de Computadores y Sistemas Operativos, Universidad Autónoma de Barcelona, España\\
\authorcr
dolores.rexachs@uab.es}

\affil[3]{III-LIDI, Facultad de Inform\'atica,
Universidad Nacional de La Plata - CIC\\
La Plata, Buenos Aires, Argentina\\
 \authorcr
erucci@lidi.info.unlp.edu.ar}

\date{{October 27, 2023}}

\begin{document}

\maketitle              % typeset the header of the contribution

\begin{abstract}
Nowadays, improving the energy efficiency of high-performance computing (HPC) systems is one of the main drivers in scientific and technological research. As large-scale HPC systems require some fault-tolerant method, the opportunities to reduce energy consumption should be explored. In particular, rollback-recovery methods using uncoordinated checkpoints prevent all processes from re-executing when a failure occurs. In this context, it is possible to take actions to reduce the energy consumption of the nodes whose processes do not re-execute. This work is an extension of a previous one, in which we proposed a series of strategies to manage energy consumption at failure-time. In this work, we have enriched our simulator and the experimentation by including non-blocking communications (with and without system buffering) and  a largest number of candidate processes to be analyzed. We have called the latter as \textit{cascade analysis}, because it includes processes that gets blocked by communication indirectly with the failed process. The simulations show that the savings were negligible in the worst case, but in some scenarios, it was possible to achieve significant ones; the maximum saving achieved was 90\% in a time interval of 16 minutes.  As a result, we show the feasibility of improving energy efficiency in HPC systems in the presence of a failure.

\blfootnote{This is the accepted version of the manuscript that was sent to review to Journal of Parallel and Distributed Computing (ISSN 1096-0848). This manuscript was finally accepted for publication on October 27hs, 2023 and its final published version is available online at \url{https://doi.org/10.1016/j.jpdc.2023.104797}.
\copyright  2023. This manuscript version is made available under the \href{https://creativecommons.org/licenses/by-nc-nd/4.0/}{CC-BY-NC-ND 4.0 license}}

\end{abstract}




\input{secciones/introduccion.tex}
\input{secciones/relatedWork.tex}

\input{secciones/background.tex}

\input{secciones/definicionEstrategias.tex}

\input{secciones/energyModel.tex}

\input{secciones/simulador.tex}

\input{secciones/experimentacion.tex}

\input{secciones/conclusiones.tex}

\section*{Funding}
This research has been supported by the Agencia Estatal de Investigacion (AEI), Spain and the Fondo Europeo de Desarrollo Regional (FEDER) UE, under contract PID2020-112496GB-I00 and partially funded by the Fundacion Escuelas Universitarias Gimbernat (EUG).

 \bibliographystyle{ieeetr} 
 \bibliography{citas}





% \end{thebibliography}
\end{document}

%% file: secciones/introduccion.tex
\section{Introduction}
Nowadays, improving the energy efficiency of high-performance computing (HPC) systems is one of the main drivers in scientific and technological research. In this context, exploring different ways to reduce energy consumption during the execution of large-scale applications is essential to maintain and increase the enormous computing power achieved. Even more, increasing the number of processing units also requires scalable fault tolerance (FT) methods. Thus, it is relevant to evaluate the energy-saving opportunities presented by these methods.

A message-passing application can be affected by failures on multiple components in distributed memory computing systems. In HPC, some methods allow continuing with the execution of the application in the presence of a failure. One of the most widely used methods is rollback-recovery through the use of checkpoints. When a failure occurs, the application can restart its execution from the checkpoints. Checkpoints can be performed in coordinated (synchronously) or uncoordinated (asynchronously) manner. In the first case, all application processes must stop, perform the checkpoint, and then continue with their execution. When a failure occurs, all processes restart from the last checkpoint. In the second case (uncoordinated checkpoints), the processes usually perform their checkpoint at different times. When a node fails, the processes of the non-failed nodes can continue their execution. As it will be shown, the uncoordinated FT method presents an interesting opportunity to explore power-saving scenarios.

The challenge is to investigate what possibilities exist to reduce energy consumption when one or more processes stop their execution due to communication blocks with processes directly or indirectly affected by the failure. From an energy-saving point of view, how to take advantage of the fact that uncoordinated checkpoints prevent all the application processes from having to go back in the presence of a failure? One possibility is to use the nodes of the processes that did not fail to execute another application of the system job queue, to use those resources for the computation of another application. Another possibility consists of keeping the affected application running and applying a series of strategies on the nodes whose processes do not need to be recovered. However, these processes, at a certain moment of execution, may be affected by long waits due to communications with processes that are recovering. Hence, if waiting is unavoidable, what is the best strategy to consume less energy at that time? 

Our objective is to know and manage the energy consumption of an HPC system by applying different strategies. The strategies to be applied depend on the underlying hardware features and the state of the application when a permanent failure occurs\footnote{Permanent failures are those which cause fail-stops in Message Passing Interface (MPI).}. Strategies that increase application execution time (in relation to the reference execution time when a failure occurs) are discarded. We consider the use of Dynamic Voltage and Frequency Scaling (DVFS) techniques and system hibernation at the node level, using the Advanced Configuration and Power Interface (ACPI). By having a characterization of the energy consumption required to execute the application and its communication pattern, it is possible to estimate the energy consumption under certain strategies from the moment of a fault. Then, by using a simulator that we have designed and developed, we can evaluate the use of the strategies. Using a simulator allows us to simplify a real system, reduce costs, and focus on essential features. In our case, it also allows us to have a flexible environment to experiment with different configurations. In addition, it integrates with a pre-existing tool that allows us to view its results.

In \cite{moran2019prediction} we show how the power consumption of checkpoint and restart operations varies by lowering the processor clock frequency.  This work, which is the continuation of \cite{moran2020towards} and \cite{moran2022some}, presents a revision of the energy model, and the following contributions:

 \begin{itemize}
   \item The definition of a series of strategies for energy saving when a failure occurs; in particular, these strategies are applied to nodes of the processes that do not have to rollback. In \cite{moran2020towards} we analyzed processes/nodes that communicate directly with the node where the failure occurred. In this work, we increase the number of candidate processes to be analyzed to apply some energy-saving strategy.
	
   \item The design and development of a simulator oriented to evaluate the proposed strategies and to select the most convenient one from the energy point of view. In this work, we extended the simulator to include non-blocking communication operations (with and without system buffering) and the analysis of the new candidate processes indicated in the previous contribution.
	
\end{itemize}

This work is organized as follows: Section~\ref{sec:relatedWork} discusses some related works. Next, Section~\ref{sec:background} presents some preliminary concepts used in the article and Section~\ref{sec:politicas} describes the strategies evaluation and application. Then, Section~\ref{sec:modeloEnergetico} presents the energy model and Section~\ref{sec:simulacion} describes the simulator. Finally, the experimental results are shown and discussed in section~\ref{sec:experimentacion}, while conclusions and future work are summarized in section~\ref{sec:conclusiones}.

%% file: secciones/relatedWork.tex
\section{Related Work}\label{sec:relatedWork}
In recent years the number of papers about energy consumption in HPC systems has been increasing. We mention here some of them, classified according to whether they are related to Fault Tolerance or not. 
 
\subsection{\textbf{Fault Tolerance and energy consumption}}

Some works evaluate the energy behavior of fault tolerance methods (rollback recovery, replication). In \cite{Saito2013}, the authors evaluate the consumption of checkpoint and restart operations at different processor clock frequencies and on different input/output devices, and design a runtime software to minimize energy consumption. In \cite{mills2014shadow} propose a computational model, called shadow computing, which provides dynamic adaptive resilience. They use redundancy to implement fault tolerance, through shadow processes, which run at a lower clock rate than the main process. They present an energy model to estimate the execution speed that minimizes energy consumption. \cite{Mills_2013} use DVFS to throttle CPU speed during checkpoint writes to measure the energy savings and performance impact. A comparative evaluation of energy consumption in three different rollback-recovery protocols (checkpoint/restart, message logging, and parallel recovery) is done in \cite{Meneses2014}. They evaluate the three protocols in a cluster with the capacity to measure the power dissipated, and develop an energy model to make projections for large-scale systems under different conditions.

A calibrator that allows measuring the energy consumption of the operations used in three methods of fault tolerance is presented in \cite{Diouri2013}. These data feed a framework (called Ecofit), to estimate the energy consumption of an application with a given TF method. Another work that also proposes a framework (called PowerCheck) that seeks to minimize energy consumption (but only for checkpoint operations) is the work presented at \cite{Rajachandrasekar2015}. The framework consists of a user-space library, which among other tasks interacts with Running Average Power Limiting (RAPL) to measure and limit the power dissipated during the checkpoint.

Some works try to estimate the optimal checkpoint interval in energy terms, like \cite{Amrizal2017} \cite{Dauwe2017} \cite{El-Sayed2015}. 

In \cite{bouteiller2013multi}, the authors propose using non-recovering nodes to execute another application from the system queue, to take advantage of the computing power of those nodes and improve the global performance of the cluster. Unlike this proposal, we also propose other strategies that use DVFS and hibernation on nodes that continue to run, without changing the application.

Finally, \cite{dichev2018energy} is the most similar proposal to this work, since they propose a localized rollback based on the data flow, and reduce the clock frequency of the waiting processes to the minimum possible. We evaluate other strategies, in addition to changing to the minimum frequency, and we do so both for the computation and waits of the processes that continue to execute.

\subsection{\textbf{Energy saving}}

This work has similarities to others that slow down the non-critical path to consume less power without substantially increasing execution time: \cite{bhalachandra2017adaptive}, \cite{rountree2009adagio}. While most of these proposals apply clock frequency adjustments during the application running, we apply it at failure time. For example, \cite{bhalachandra2017adaptive} uses a per-core Software Controlled Clock Modulation or DVFS to throttle the frequencies of cores not on the critical path of an MPI application. In \cite{rountree2009adagio} they adjust the clock frequencies   of   the tasks in a manner that they arrive just in time at the synchronization moment.
 
%\subsection{\textbf{contadores de hw}} 
Some works use hardware counters to predict energy consumption, f.e. \cite{wu2020performance}, where they run various machine learning methods, measure hardware counters, and analyze which ones correlate best with runtime, and  CPU   and memory power consumption. To do this, the authors get some functions to predict the time and energy savings when applying possible optimizations to the applications. In \cite{wang2011span} they propose a model for estimating the power consumed by an application using the least amount of performance monitoring counters possible.

%\subsection{\textbf{Similar works}} 
In \cite{dolz2012simulator}, they perform a simulation of an HPC workload to evaluate the energy savings when activating and deactivating nodes according to the computational and power requirements of the cluster. In this work, we also shut down nodes when the waiting phase is long enough. Other works analyze the active waits of an MPI application and evaluate potential energy savings by changing the clock frequency during those waits, as in \cite{knobloch2012determine}, or entering a power-gated state executing a sleep call, as in \cite{piga2016performance}. Active waits are also an energy-saving point in our proposal.

%% file: secciones/background.tex
\section{Background}\label{sec:background}
The following subsections introduce some concepts that are used in the article, such as the states defined by ACPI, operation modes and active waits in MPI, and rollback recovery mechanism. 

\subsection{ACPI}\label{sub:backgroundACPI}
The ACPI specification provides an open standard that allows the operating system to manage the power of the computing system and provides advanced mechanisms for energy management\footnote{https://www.uefi.org/specifications}. The specification defines a series of global states and substates for the system. In the global \textit{working} G0 state the applications are executed. In this global state, there is one executing state (C0) and some inactive states (C1..Cx). Inside C0, performance states (P states) are defined, which determine performance and power demand. In the global \textit{sleeping} state G1 the computer consumes a small amount of power and applications are not executed. In this state, the context is saved, so the operating system does not need to restart when waking up. The latency for returning to the working state varies on the type of sleeping substates selected (S1-S4). There are two more states, \textit{soft} and \textit{mechanical off} not used in this work. Table~\ref{tab:estadosACPI} shows a brief description of the states and substates.

In this work, we use performance and sleeping states (P and S states). In GNU/Linux, it is possible to change the frequency of the processor (P states) with the  userspace governor, which permits modifying the frequency of each core. To change the sleeping state (S states) a string has to be written to the $/sys/power/state$ file. In the simulations, when we mention the strategy of sleeping the node we refer to the Suspend-to-RAM (S3) state, and when we mention clock frequency changes, the same change is applied to all cores of the node.

\begin{table}[htbp]
	\caption{ACPI global states and substates}
	\begin{tabular}{|p{4cm}|p{9cm}|}
		\hline
		Global State      & Substates       \\
		\hline
		G0 Working  & Processor Power States. \textbf{C0}: Executing state.  A processor that is in the C0 state will also be in a Performance State (Px states). The \textbf{P0} state means an execution at the maximum capacity of performance and power demand. As the number of P state increases, their performance and demanded power is reduced. The processors implement the P states using the technique of DVFS. \\
		&   \textbf{C1...Cx}: inactive states. Very short transition latencies in comparison to transition latencies between states G0-G1\\
		\hline
		G1 Sleeping   & \textbf{S1-S4} are idle states of the system associated with G1. Increasing the state number requires less power and longer latency to exit the state. There are no transitions between S1-S4, it is always with S0.\\ 
		\hline
		G2 Soft Off & \textbf{S5}. It consumes a minimum of energy. No applications are running and the system context is not saved. \\
		\hline
		G3 Mechanical Off & The system is completely off and there is no electrical current running through the circuit.\\
		\hline
	\end{tabular}
	\label{tab:estadosACPI}
\end{table}

\subsection{Communication operations in MPI}\label{sub:operationsMPI}
The MPI standard provides blocking and non-blocking operations for message passing \footnote{https://www.mpi-forum.org/docs/mpi-4.0/mpi40-report.pdf}. Blocking send operations are characterized by returning as soon as it is safe to reuse the buffer (either because the message was copied into the receiving buffer or a sender's system buffer). In standard mode (MPI\_Send, MPI\_Recv), it is up to the MPI implementation to decide whether outgoing messages will be buffered. If the system provides buffering, the send call may complete before a matching receive is invoked. If the system does not provide buffering, the send call will not complete until a matching receive operation has been posted, and the data has been moved to the receiver buffer.

\begin{figure*}
	\centering
	\includegraphics[width=\columnwidth]{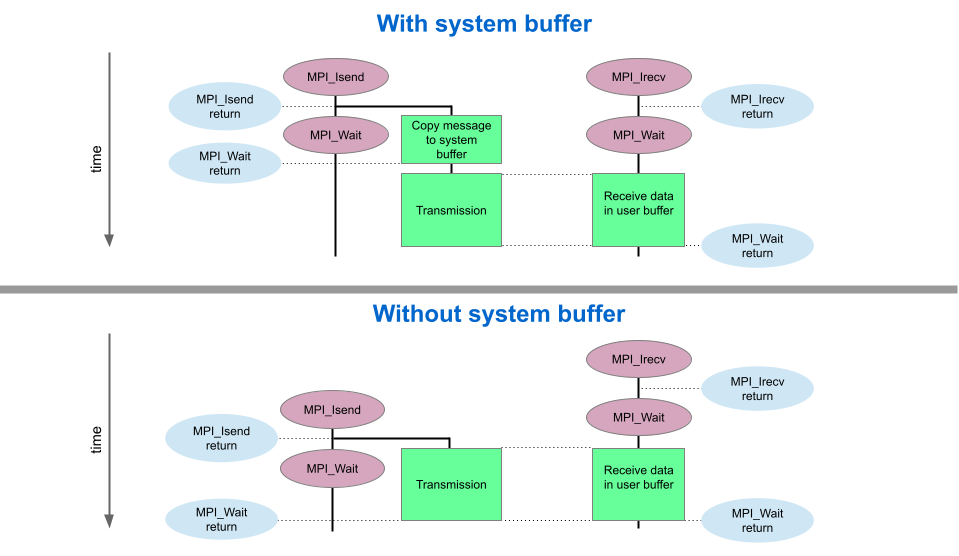}
	\caption{Non-blocking operations return point.}
	\label{fig:punto_retorno}
\end{figure*}

Non-blocking operations return immediately, and another operation, usually called wait (MPI\_Wait), is used to verify that it is safe to reuse the buffer for the send case, or that the data is already in the receive buffer for the receive case. This allows us to overlap computation with communication. In the standard mode (MPI\_Isend, MPI\_Irecv), if the system buffer is used, the wait will return after copying the data to that buffer. If no system buffer is present, the wait will return after the data is copied to the receiver buffer. Fig.~\ref{fig:punto_retorno} shows the return point of the non-blocking operations with and without using the system buffer. In this work, we consider both blocking and non-blocking operations in their standard mode.

\subsection{Active and idle waiting in MPI}\label{sub:esperasMPI}
In MPI parallel applications, it may happen that one process must wait for another to send or receive a message. During these waits, the process can keep the processor busy by active-waiting, or releasing it, and using polling or interrupts. Active-waiting means polling at the highest available frequency for a signal to be able to react instantly once the signal is received \cite{knobloch2012determine}. An active wait keeps the processor busy and consumes energy without doing useful work. An idle wait can affect application performance, due to C states transitions \cite{cesarini2018countdown}; this is why various MPI implementations provide active wait as the default operating mode. As this operating mode is configurable, in this work we consider both cases.

\subsection{Rollback recovery}\label{sub:rollback}

A consistent global state can be found during a successful and fault-free execution of parallel computing. Inconsistent states can occur because of failures. A fundamental goal of any rollback-recovery protocol is to lead the system to a consistent state after a failure. This method consists of periodically saving the state of the application in stable storage, which is known as a \textit{checkpoint}. At failure time, it is possible to restart the application from the last successfully saved state, which is called \textit{restart}. In the case of coordinated checkpoints, a consistent global state is obtained by synchronizing all the processes, stopping their execution, and performing the checkpoint. When a process fails, all processes must restart from the last checkpoint. As we can see, all the application processes re-executing produce energy and time overhead. In the case of uncoordinated checkpoints, processes take their checkpoints independently, avoiding synchronization time and I/O contention~\cite{levy2014using}. At failure time, only failed processes restart from the last checkpoint, using fewer resources for their recovery than a coordinated checkpoint. However, ensuring a consistent global state is not as straightforward as in the case of coordinated checkpoints. When a process restarts, orphaned and/or lost messages can appear, causing other processes to roll back to ensure consistency. This is called \textit{domino effect}, and there are different techniques to control it, such as the use of message logging \cite{meyer2017hybrid}. Uncoordinated checkpoints allow the use of advanced checkpoints. If a process is going to block by communication, and its last checkpoint happened a relatively long time ago, the process performs a checkpoint before blocking. In this way, useless waiting time is used by a checkpoint operation. In this work, we consider the use of advanced checkpoints.

Hybrid approaches also exist and they take advantage of coordinated and uncoordinated checkpoints. In this scheme, the processes are divided into groups. Within each group, a coordinated checkpoint scheme is used, but between groups, an uncoordinated checkpoint strategy is followed. There are different criteria for defining groups. For example, all processes running on the same node could be in a group, because when a node fails all its processes must restart\cite{castro2015fault}. Another way to define groups can be with processes that communicate frequently \cite{ho2008scalable}. The first approach is the one used in the present work.

%% file: secciones/definicionEstrategias.tex
\section{Strategies definition and application}\label{sec:politicas}

The use of uncoordinated checkpoints as a fault tolerance method of a message passing HPC application enables one to take actions on the nodes that should not recover from the failure. At the time of a failure, it is possible to apply two different options, which depending on the state of the system, will have a different impact on indices such as cluster productivity and energy consumption.
One option is to change the application and the other one is to maintain it, but changing  the P and S states of the surviving nodes. Both options are defined and analyzed below.

\subsection{\textbf{Change of the application}}

When a node fails, one possible action to take is to change the application, as proposed in \cite{bouteiller2013multi}. One of the nodes that did not fail is used for process recovery (restart and re-execution). The rest of the nodes are used to run another application in the job queue. In this way, the productivity of the computer system is increased. To apply this option, the recovery time must be long enough so that it is worth changing the application. While the processes that were running on the failed node recover, the rest of the nodes do useful work allowing another application to move forward.

Switching between applications is a costly task in terms of time and resources because the filesystem has to be accessed multiple times, with consequent energy consumption. The checkpoints are performed on the application that comes out (write operations), the checkpoints of the incoming application are loaded (read operations) and the restart is executed (read operations). These steps must then be repeated to reload the original application. If the re-execution time is long enough to outweigh the cost of these operations, this alternative allows nodes that did not fail to continue running at maximum performance. Allowing another application to proceed while another recovers from a failure improves the overall productivity of the system.

\subsection{\textbf{Changes of the P and S states}}\label{subsec:cambiosPyS}
This proposal focuses on energy saving and that is why a set of strategies based on changing the P and S states (see subsection~\ref{sub:backgroundACPI}) is evaluated to apply to the nodes that did not fail. This subsection defines the possible strategies to apply after the failure of a node when the application is not changed. In addition, it shows why it is important to consider both direct and indirect process blocking, and discusses some situations to take into account in the application of the strategies.

\begin{figure*}[t]
	\centering
	\includegraphics[width=\textwidth]{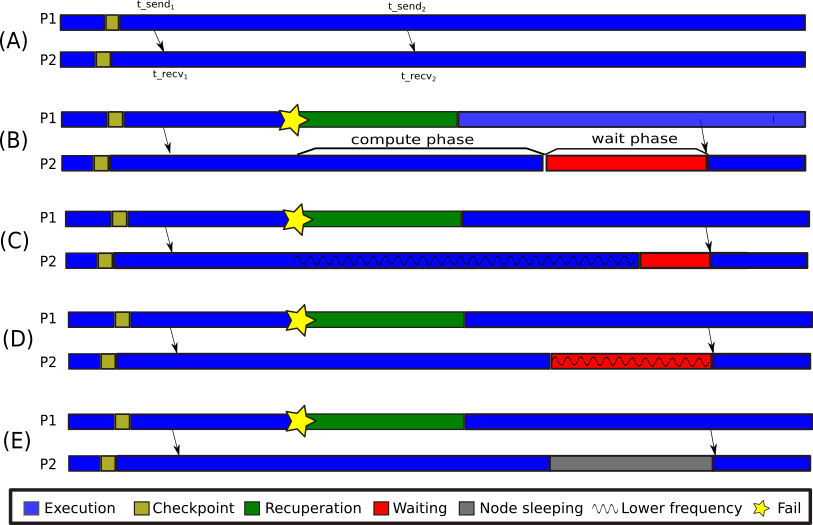}
	\caption{Strategies application for different failure cases. (A) No failure occurs. (B) A failure occurs and no action is taken (reference case). (C)  A failure occurs and frequency change for the computational phase is taken. (D)  A failure occurs and frequency change in the waiting phase is taken. (E)  A failure occurs and sleeping in the waiting phase is taken.}
	\label{fig:casos}
\end{figure*}

Fig.~\ref{fig:casos} shows different scenarios, where two processes, P1 and P2, are running on different nodes. The little squares indicate that the process is performing checkpoints. As uncoordinated checkpoints are used, they may be performed at different times, as shown in the figure. \textbf{Case A} shows a failure-free execution, where P1 sends two messages to P2, indicated by $t\_send_1$ and $t\_send_2$. Messages are sent and received in an order scheduled by the application, which maximizes the time dedicated to useful computation while minimizing the waits for communications (data transfers and synchronizations). In an efficient parallel application, waits not related to the failure are minimal. For simplicity, we consider these waits within the computation phase. \textbf{Case B} to \textbf{case E} shows an execution where process P1 fails (indicated by a yellow star) and must recover.  This figure (and the next one) assumes the sending process always fails. However, this does not affect the application of the strategies, because this application is not affected by whether the failing process is the sender or the receiver. Instead, the analysis focus on the moment in which the alive process gets blocked by communication due to a failure.  After a failure, the restart is initiated, and the re-execution begins ( indicated in blue). We can see how the sending of the second message is delayed due to the failure, and the process P2 must wait (indicated in red or gray). In \textbf{Case B} no action is taken, and it will serve as a reference for the evaluation of the strategies. In this case, the computing and waiting phases are indicated. These phases are defined for processes that do not fail, and are defined as follows:

\begin{itemize}
	\item  The \textbf{compute phase} comprises the execution of the application from the moment of failure until it gets blocked because of the failure. This blocking is caused by a communication operation with a process that has failed or that is blocked because of the failure. Communications with other processes are included in the computing phase. In this way, additional waits can appear, not related to the failure. It may also happen that during this phase checkpoints are performed. 

	\item The \textbf{waiting phase} begins with the blocking related to the failure and ends when the communication concludes and the process begins to compute again. 
	
\end{itemize}

At failure time, the best strategy for the computation and waiting phase is determined. The strategy for the computation phase can be to lower the clock frequency. The strategies for the waiting phase can be to sleep the node (in some state S1-S4) or to switch to the minimum clock frequency. The computation and waiting  phases altogether, form the \textit{intervention interval}. 

In \textbf{case C}, the P2 node changes to the selected clock frequency during the computation phase, indicated by the wavy line. This makes P2 submit the second receive operation later, shortening the waiting phase. The selected clock frequency must meet two criteria:

\begin{enumerate}
	\item The duration of the computation phase running with the selected frequency must not exceed the synchronization time of the process. That is, the selected frequency must not cause waits in other processes.
	
	\item The energy consumption of the compute phase in conjunction with the waiting phase should be lower than the consumption in the reference case (case B).
\end{enumerate}

The selected frequency may be the maximum frequency, in which case there is no change in frequency. For simplicity, we always select a single frequency for the entire computation phase, although it could be the case that it is a combination of two or more frequencies that best meet the criteria just mentioned (applying each frequency during a fraction of the computation phase).

In \textbf{case D}, a change of the clock frequency during the waiting phase was decided, indicated by the wavy line. In \textbf{case E}, the node was sent to sleep during the waiting phase. Neither of these actions affects the waiting phase duration, but it does impact energy consumption indeed.

The strategies explained above can be applied in combination. For example, it could be the case that the clock frequency is changed for the computation phase and the waiting phase. As we can see, there are several possible scenarios where different actions must be evaluated and managed.

The strategies can be summarized as follows:
\begin{itemize}
	\item Frequency change for the computational phase (case C). 
	\item Frequency change for the waiting phase (case D).
	\item Sleeping for the waiting phase (case E).
\end{itemize}

The evaluation of the strategies for the computing and waiting phase is done altogether, considering the impact on energy consumption and execution time. The selected configuration will be the one that achieves the lowest energy consumption for the \textit{intervention interval}, without affecting the application execution time. For this, the selected frequency for the computation phase of a process should avoid that other processes having to wait for it. Regarding the waiting phase, if the duration of this phase is long enough to sleep the node and subsequently wake up it, achieving a lower energy consumption, then this option is selected. Otherwise, if the waiting operations are configured to be active waits, the way to minimize energy consumption is by using the minimum frequency. If the waiting operations are configured to be idle waits, the way to minimize energy consumption is to do nothing \cite{knobloch2012determine}.

\subsubsection{\textbf{Cascade blocking and communications depth}}

\begin{figure*} [t]
	\centering
	\includegraphics[width=\textwidth]{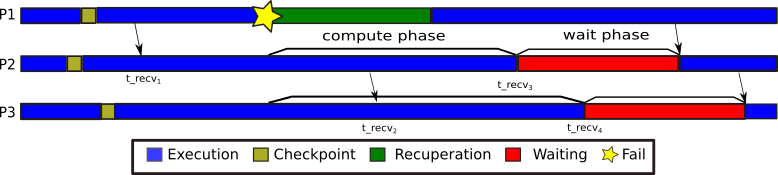}
	\caption{Cascade blocking and communications depth.}
	\label{fig:bloqueosCascada}
\end{figure*}

Now we consider the processes which are indirectly affected by a failure. In Fig.~\ref{fig:bloqueosCascada}, P3 does not block with the failed process; it blocks with an alive process, that is blocked because of the failure. We name \textit{cascade blocking} to these blocking situations, which arise as a result of the failure but are not directly related to the failed process. These blockings are propagated over a set of processes during the execution of the application from the moment of failure. This situation can occur in the immediate following communication, or in the  subsequent ones. We define \textit{depth} to be the number of subsequent communications to consider after the failure, looking for one that is blocked for this reason. In Fig.~\ref{fig:bloqueosCascada}, P3 blocks on the second communication with P2, giving a depth of 2. The depth can be calculated by looking at the pattern of communications between each pair of processes and choosing the maximum number of communications found in the pattern.

\subsubsection{\textbf{Estimation of the compute and waiting phases}}

\begin{figure*}[t]
	\centering
	\includegraphics[width=.75\columnwidth]{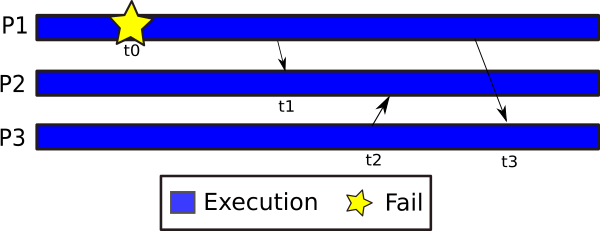}
	\caption{View of three processes at the time of failure $t0$.}
	\label{fig:ejemplo3procesos}
\end{figure*}

To choose the strategies an algorithm has been developed that estimates the blocking times of each process. In this algorithm, we call \textit{children} of a process, the processes with which it communicates. The main steps are summarized below:

\begin{itemize}
	
	\item First, the lists are filled; $List1$ contains the processes of the parent level, initially the processes of the failed node. $List2$ contains the child processes of the processes in $List1$, with the times at which they are estimated to block for communication with their parents. For example, for the case shown in Fig.~\ref{fig:ejemplo3procesos}, the lists would be as follows:
	\begin{verbatim}
		List1={P1}
		List2={(P2,t1);(P3,t3)}
	\end{verbatim}
	
	\item  The algorithm then visits each $List2$ process and checks if it will be blocked earlier by another $List2$ process (sibling process), in which case it updates the block time. Continuing with the example, by visiting the first element of $List2$, we see that P2 does not block before with any sibling process. Visiting the second element of $List2$, P3, we see that it blocks with P2 at a time before the current block time of P3, and after the block time of P2. That is, P3 blocks at a time $t2$ that verifies that $t1 < t2 < t3$. After the update, $List2$ would look like this:
	\begin{verbatim}
		List2={(P2,t1);(P3,t2)}
	\end{verbatim}

	\item Finally, we proceed to the next level, assigning $List2$ to $List1$ and emptying $List2$. After this step, the lists would be as follows:
	\begin{verbatim}
		List1={(P2,t1);(P3,t3)}
		List2={}
	\end{verbatim}
	These steps are repeated until there are no more processes affected by the failure and unanalyzed.
	
\end{itemize}

The output of the algorithm is the time at which each process is expected to block due to the failure. The complexity of this algorithm is in the order of the number of processes by the number of communications of each process. For more detail, you can see the pseudocode in Algorithm~\ref{alg:algCascada}.

\begin{algorithm}
\caption{Pseudo-code to analyze cascade-blocked processes}\label{alg:algCascada}
\begin{algorithmic}[1]
	
	\STATE List1 = {$(P_{failed},0)$};  \CommentInline {The elements of the lists are a pair (process id, block time)}
	\STATE List2 = ListGlobal = \{\}; 
	\WHILE{there are processes that communicate with processes in List1 that do not belong to ListGlobal}

		\FOR{each ($P_{parent}$, parent\_block\_time) in List1}
			\FOR{each process $P_{child}$, child of $P_{parent}$}
				\STATE comm\_time = next\_comm($P_{child}$, $P_{parent}$);  \CommentInline{Find the next communication between both processes}
				\STATE i = 0;
				\WHILE{(comm\_time $<$ parent\_block\_time) AND (i$<$depth)}
					\STATE comm\_time = next\_comm($P_{child}$, $P_{parent}$); 
					\STATE i++;
				\ENDWHILE
				\IF {$P_{child}$ belongs to List2}
					\STATE block\_time = get\_block\_time($P_{child}$);
					\IF {comm\_time $<$ block\_time}
						\STATE Update\_List2($P_{child}$, comm\_time) \CommentInline{Update $P_{child}$ block time}
					\ENDIF
				\ELSE
					\STATE List2 = List2 +  ($P_{child}$, comm\_time); \CommentInline{Add $P_{child}$ to List2}
				\ENDIF
			\ENDFOR	
		\ENDFOR
		
		\COMMENT{List2 Convergence}
		\STATE $keep\_analyzing$ = false;
		\REPEAT
			\FOR{each process $P_i$ in List2}
				\IF {exist $P_x$ in List2 that communicates with $P_i$ in time $t$ such that: \\
					$P_x\_block\_time$ $<$ t $<$ $P_i\_block\_time$}
					\STATE UpdateList2($P_i$, t); \CommentInline{Update $P_i$ block time}
					\STATE $keep\_analyzing$ = true;
				\ENDIF
			\ENDFOR
		\UNTIL  {not $keep\_analyzing$}
		\STATE ListGlobal = ListGlobal + L1; \CommentInline{group the processes already analyzed}
		\STATE List1 = List2; \CommentInline{List2 levels up}
		\STATE List2 = \{\}; \CommentInline{List2 is emptied to incorporate the processes of a new level}
	\ENDWHILE 		
	\end{algorithmic}
\end{algorithm}

This algorithm is suitable for applications with long compute phases, which allow the application of some strategies in these phases. In applications with short compute phases, where the impact of the frequency change is likely to be negligible, it may be convenient to estimate only the waiting phases and evaluate the strategies for these phases.

\subsubsection{\textbf{Evaluation of the energy-saving strategies}}

Once the computing and waiting phases of each process have been estimated, the possible strategies are evaluated. For this, the compute phase of each live process is instantiated with each of the possible clock frequencies, taking into account the restriction of not affecting the reference time. Based on this, the new duration of the waiting phase and the appropriate action to take are evaluated. Then the combination of strategies that obtain the lowest estimated energy consumption in each case is selected. In a previous work \cite{moran2020towards}, we showed a pseudocode for strategy evaluation, which has been implemented in a simulator. This algorithm implements what is described in the energy model in section~\ref{sec:modeloEnergetico}, and is in the order of the number of clock frequencies by the number of processes. The evaluation of the strategies for each process can be computed in parallel.

\subsubsection{\textbf{Secondary effects of the frequency change in the compute phase}}
When a node changes its clock frequency to a lower frequency, all the processes running on that node are slowed down. The time in which communication takes place between these processes and other non-slowed (or differently slowed) processes will be affected. This could result in an error in the estimation of the compute and waiting phases of these last processes, and a suboptimal application of the strategies. One possible solution is to evaluate processes in the same order that blocking propagation occurs, resulting in a slowdown propagation. For simplicity, the simulator does not implement it.

\subsubsection{\textbf{Secondary effects of sending a node to sleep}}
When using non-blocking MPI operations (see Subsection~\ref{sub:operationsMPI}), sending to sleep a node may affect application execution. For example, consider the following situation. A process issues a non-blocking $send$, and then its node is sent to sleep by the application of the strategy. Later, another process emits the corresponding $receive$, being not able to receive the data that is in the buffer of the sending process because the node of that process is sleeping. This causes an unexpected wait because if the strategy had not been applied, the node would be awake and the communication would have been done. One possibility is, in these cases, not sending the node to sleep. To do this, it would be necessary to keep a registry of the pending non-blocking communications. Another option could be to wake up the node when required. In the latter case, if active waits are used (see Subsection~\ref{sub:esperasMPI}), and the remaining wait time justifies it, the clock frequency should be changed to the minimum available. That is, there would be a change of strategy in the middle of a waiting phase.

%% file: secciones/energyModel.tex
\section{Energy model}\label{sec:modeloEnergetico}
The model presented below estimates the energy savings obtained when the best strategy is applied. The input of the model is application, fault tolerance and system data, and the estimated duration of the compute and waiting phases of each process, as indicated in Table~\ref{tab:entradas_modelo}. To calculate the energy consumption during an interval of time we need to know the interval duration and the associated average power dissipation. Power and time can be obtained from characterizations as in \cite {moran2019prediction}. For simplicity, we consider the computation to be homogeneous, i.e. the application dissipates the same power throughout its entire execution. The application communication pattern can be obtained from the execution trace \cite{wong2010pas2p}. The duration of the compute and wait phases (see subsection~\ref{subsec:cambiosPyS}) have to be estimated at failure time. Table~\ref{tab:parametros_modelo} details model parameters for reference. Case B of Fig.~\ref{fig:casos} reflects the situation where no strategies are applied and serves as a reference.

\begin{table}
\centering
\caption{Energy model inputs}
\begin{tabular}{ |p{4cm}|p{8cm}|  }
\hline
System data &  Power and time required to sleep and to wake up a node, downtime. \\
\hline
Application data & Power dissipated and slowdown factor for each frequency during the computation. Pattern and frequency of communication among processes. \\
\hline
Fault tolerance data & Checkpoint and restart duration. The power dissipated and the slowdown of each frequency during checkpoint\\
\hline
Variables &  Compute and wait phase duration, number of checkpoints during intervention interval.\\
\hline
\end{tabular}
\label{tab:entradas_modelo}
\end{table}

The energy saving obtained with the application of the strategies is estimated as the sum of the savings in each node, as shown in Eq. (\ref{eq:ahorroTotal}), where $n$ is the number of nodes where the application is executed. 

\begin{equation}
Total\_Energy\_Saving = \sum_{j=1}^{n} Energy\_Saving_{node\_j}
\label{eq:ahorroTotal}
\end{equation}

The energy savings of each node (Eq.~\ref{eq:ahorro}) is calculated as the difference between the estimated energy consumption without intervention, $ENI$ (\underline{E}nergy \underline{N}o \underline{I}ntervention) and the minimum estimated energy consumption with intervention, $EI$ (\underline{E}nergy with \underline{I}ntervention).

\begin{equation}
Energy\_Saving_{node\_j} = ENI(f_{max}) - Min\{EI(f_i)\}
\label{eq:ahorro}
\end{equation}

The energy consumption without intervention, $ENI$, is shown in Eq.~\ref{eq:ENI}. This equation is instantiated with the maximum available clock frequency.

\begin{equation}
ENI(f_{max})= E\_comp(f_{max}) + E\_awake\_wait(f_{max})
\label{eq:ENI}
\end{equation}

The energy consumption with intervention, $EI$, is shown in Eq. ~\ref{eq:EI}. This equation is instantiated with each of the clock frequencies, $f_i$. Once we have all the estimations using all the frequencies, the lowest one is selected (as indicated in Eq.~\ref{eq:ahorro}).

\begin{equation}
EI(f_i)= E\_comp(f_i) + EI\_wait(f_i)
\label{eq:EI}
\end{equation}

In Eq. ~\ref{eq:ENI} and~\ref{eq:EI}, the energy consumed by the node is estimated as the sum of the energy consumed during the computation phase plus the energy consumed during the waiting phase.

In all the equations, the clock frequency (indicated as an argument) refers to the frequency at which the computation phase will be executed. In particular, in the equations referring to the waiting phase, this argument is necessary to estimate the duration of this phase (when the computation phase is executed with that frequency). Different frequencies in the computation phase can cause different actions to be applied in the waiting phase.

\subsection{\textbf{Compute phase energy}}

To estimate the energy consumed during the computation phase we use Eq. ~\ref{eq:Ecomp}. 

\begin{equation}
\begin{aligned}
E\_comp(f_i)= T\_comp(f_i) \times P\_comp(f_i) + \\
N\_ckpt \times T\_ckpt(f_i) \times P\_ckpt(f_i)
\end{aligned}
\label{eq:Ecomp}
\end{equation}

In this equation, the duration of the compute phase is calculated for the selected frequency using the slowdown factor $beta$, as indicated in Eq.~\ref{eq:tcomp}.

\begin{equation}
T\_comp(f_i) = T\_comp(f_{max}) \times \beta(f_i)
\label{eq:tcomp}
\end{equation}

The slowdown factor can be characterized as in Table~\ref{tab:conf1} and indicates how much slower the application runs with frequencies lower than the maximum one. The duration of the possible checkpoints that are performed in this stage is calculated in the same way (Eq.~\ref{eq:tcheck}).

\begin{equation}
T\_check(f_i) = check\_time \times \gamma(f_i)
\label{eq:tcheck}
\end{equation}

The number of checkpoints that are carried out during the intervention interval (usually one or none) is an input to the model. The power dissipated with each frequency, $P\_comp(f_i)$, is an input to the model.

\subsection{\textbf{Waiting phase energy}}

To estimate the energy consumed during the waiting phase without intervention ($ENI$) we use Eq.~\ref{eq:E_awake_wait},~\ref{eq:E_active_wait} and~\ref{eq:E_idle_wait}.

\begin{equation}
E\_awake\_wait(f_i) =
\begin{cases}
E\_active\_wait(f_i) & \text{  if active wait}  \\
E\_idle\_wait(f_i) & \text{  if idle wait}
\label{eq:E_awake_wait}
\end{cases}
\end{equation}

\begin{equation}
E\_{active\_wait}(f_i) = T\_wait(f_i) \times P\_active\_wait(f_i)
\label{eq:E_active_wait}
\end{equation}

\begin{equation}
E\_{idle\_wait}(f_i) = T\_wait(f_i) \times P\_idle\_wait
\label{eq:E_idle_wait}
\end{equation}\\

In Eq.~\ref{eq:E_awake_wait}, the energy consumption will be determined by the message waiting configuration. If the configuration indicates that active waits are used (Eq.~\ref{eq:E_active_wait}), the energy consumed is calculated using the power dissipated by the corresponding frequency. On the other hand, if idle waits are used (Eq.~\ref{eq:E_idle_wait}), the processor is practically idle, and the energy consumed is calculated using a power that is close to the \textit{base power}. We call \textit{base power} the power dissipated when the node is in Working G0 state, i.e. no jobs are running and most of the cores are in inactive states (see Table~\ref{tab:estadosACPI}).

To estimate the energy consumed during the waiting phase with intervention ($EI$) we use Eq.~\ref{eq:EI_wait}. This equation distinguishes two cases, depending on whether the node goes to sleep or stays awake.

\begin{equation}
EI\_wait(f_i) =
\begin{cases}
E\_awake\_wait(f_i) & \text{ if no sleeping} \\
E\_sleep\_wait(f_i) & \text{ if sleeping} 
\label{eq:EI_wait}
\end{cases}
\end{equation}

If the node stays awake we use the Eq.~\ref{eq:E_awake_wait},~\ref{eq:E_active_wait} and~\ref{eq:E_idle_wait} described above. If the node goes to sleep, the Eq.~\ref{eq:E_SleepWait} is used. This equation estimates the energy considering the power dissipated during the transitions between the active-sleep-active states and while the node is sleeping.

\begin{equation}
\begin{aligned}
EI\_sleep\_wait(f_i)=  T\_go\_sleep \times P\_go\_sleep + \\
T\_sleep(f_i) \times P\_sleep + \\ 
T\_wakeup \times P\_wakeup
\end{aligned}
\label{eq:E_SleepWait}
\end{equation}

The energy consumed to sleep and wake up the node will depend on the state of the node. For simplicity, we use a single value. To sleep a node, two conditions must be met. First, the waiting time must be greater by a certain factor, $\mu_1$, than the total time that the node requires to sleep and wake up. This is to prevent sleeping a node for a period shorter or equal to the time it takes to put it to sleep and to wake it up. This condition is expressed in Eq.~\ref{eq:cond1}.

\begin{equation}
\begin{aligned}
T\_wait(f_i) > \mu_1 \times (T\_go\_sleep + T\_wakeup) \end{aligned}
\label{eq:cond1}
\end{equation}

Secondly, the energy consumption while sleeping (including the energy consumed going to sleep and waking up) must be lower, by a certain factor $\mu_2$, than the consumption obtained if the node remains awake. This is to avoid sleeping a node that will have almost the same power savings as staying awake. This condition is expressed in Eq.~\ref{eq:cond2}.

\begin{equation}
\begin{aligned}
E\_sleep\_wait(f_i)  < \mu_2 \times E\_awake\_wait(f_{min}) 
\end{aligned}
\label{eq:cond2}
\end{equation}

\begin{table}[htbp]
\caption{Parameters}
\begin{center}
\begin{tabular}{|p{4cm}|p{8cm}|}
%\begin{tabular}{|l|l|}
\hline
Parameter Name & Description \\
\hline
$E\_comp(f_i)$ & Energy consumed by the node during the computing phase, at frequency $f_i$. \\
\hline
$E\_awake\_wait(f_i)$     & Energy consumed by the node during the waiting phase when it remains awake.\\
\hline
$EI\_wait(f_i)$     & Energy consumed by the node during the waiting phase when the intervention takes place.\\
\hline
$E\_active\_wait(f_i)$ & Energy consumed by the node during the active waiting phase at frequency $f_i$.\\
\hline
$E\_idle\_wait(f_i)$ & Energy consumed by the node during the waiting phase when using idle wait.\\
\hline
$E\_sleep\_wait(f_i)$ & Energy consumed by the node during the waiting phase when it goes to sleep. \\
\hline
$T\_ckpt(f_i)$ & Checkpoint duration running at frequency $f_i$.\\
\hline
$T\_comp(f_i)$ & Computation phase duration when node executes at the frequency $f_i$.\\
\hline
$T\_go\_sleep$ $T\_wakeup$ & Times required by a node to sleep and wake up, respectively.\\
\hline
%$T\_{failed_{i}}$ & Time required by a failed process to recover and to block with a process of node $i$.\\
%\hline
%$T\_recover$ & Time required by a failed process to recover and return to the point where the failure occurred. \\
%\hline
%$T\_down$ & Downtime. \\
%\hline
%$T\_rest$ & Restart duration at maximum frequency. \\
%\hline
%$T\_reexec$ & re-execution time at maximum frequency. \\
%\hline
$T\_sleep(f_i)$ & Time that node is sleeping when compute phase is executed at frequency $f_i$, without considering the time to go to sleep and to wake up.\\
\hline
$T\_wait(f_i)$ & Waiting phase duration when computation  phase is executed at frequency $f_i$. \\
\hline
$T\_ckpt$ & Checkpoint duration. \\
\hline
$P\_go\_sleep$ $P\_wakeup$ & Power dissipated while sleeping and waking up a node, respectively.  \\
\hline
$P\_sleep$ & Power dissipated when the node is sleeping.  \\
\hline
$P\_comp(f_i)$ & Power dissipated during computation when running at frequency $f_i$.\\
\hline
$P\_ckpt(f_i)$ & Power dissipated during checkpoint at frequency $f_i$.\\
\hline
$P\_active\_wait(f_i)$ & Power dissipated during active wait at frecuency $f_i$. \\
\hline
$P\_idle\_wait$ & Power dissipated during the idle wait. \\
\hline
$N\_ckpt$ & Number of checkpoints during the computation phase (zero or one).\\
\hline
%$\alpha_{ij}$ & Percentage of the communication interval between process $i$ and process $j$ that still remains to be executed for process $i$ to block in a communication with process $j$.\\
%\hline
%$I\_comm_{ij}$ & Duration of the communication interval between process $i$ and process $j$ when executing at maximum frequency.\\
%\hline
$\beta(f_i)$ $\gamma(f_i)$ & Slowdown of instruction and checkpoint execution when frequency $f_i$ is used. \\
\hline
$\mu_1$ $\mu_2$ & Time and energy threshold to determine whether or not to sleep a node. \\
\hline
\end{tabular}
\label{tab:parametros_modelo}
\end{center}
\end{table}

%% file: secciones/simulador.tex
\section{The simulator}\label{sec:simulacion}
We have developed an event-based simulator that uses the SMPL library \cite{macdougall1987simulating} written in C language. This simulator allows us to evaluate the strategies under different system configurations, different characteristics of the application and different failure times. The main features of the simulator are detailed below:
\begin{itemize}
	
	\item The failure of a node in a parallel message-passing application with uncoordinated checkpoints at the system level is simulated. For simplicity, a single process per node is simulated; in particular the most representative process of the node. 
	
	\item We refer to the representative process of a node as  the one that first blocks due to the failure. This simplification is based on the fact that if the processes are assigned to a node by affinity, the communication blocks of all the processes in the node will be at a similar time to that of the simulated process. 
	
	\item In the same line as the previous point, the strategy selected when evaluating the representative process is applied to the node. 
	
	\item Checkpoints can be triggered by events or by time; as we seek to simulate transparent checkpoint to the application, we activate it by time. 
	
	\item The moving ahead of checkpoints is simulated (see subsection~\ref{sub:rollback}). 
	
	\item The checkpoint files are stored in a parallel file system external to the nodes. 
	
	\item The message log and the domino effect have not been considered.
	
	\item The messages have a fixed size. 
	
	\item The overhead caused by the strategies evaluation and implementation is not considered. 
	
	\item The simulated MPI functions are blocking and non-blocking standard mode (see subsection~\ref{sub:operationsMPI}). 
\end{itemize}

The simulator inputs are the same as the energy model described in subsection~\ref{sec:modeloEnergetico}, and are detailed in Table~\ref{tab:entradas_modelo}. At the time of failure, the simulator evaluates the model for each surviving process with each of the clock frequencies provided, and determines the best strategy to apply. The simulator output includes the estimated energy savings when applying the selected strategy, and a trace to visualize the behavior of the application execution. The trace is visualized with the Paraver\footnote{http://www.bsc.es/computer-sciences/performance-tools/paraver} tool, a flexible HPC application performance analysis and visualization tool.

%% file: secciones/experimentacion.tex
\section{Experimentation and results analysis}\label{sec:experimentacion}

The following subsections describe the experimental work carried out with the simulator. The proposed strategies seek to exploit different opportunities to reduce the energy consumption of the applications affected by the failures. The experiments present specific scenarios to show the validity of the proposal. The results obtained are analyzed and
some discussions are at the end of the section.

\subsection{Experimental settings}

Table~\ref{tab:conf1} shows dissipated power and slowdown factor ($\beta$ and $\gamma$) obtained from measurements on a six-core Intel Xeon E5-2630 node, with a frequency range of 1.2 GHz to 2.8 GHz (with the mechanism Intel Turbo Boost disabled). The base power is 60W. The node sleep and wake times are set at 25 and 5 seconds respectively, and the average powers at 51 and 91 watts respectively. The average power dissipated while the node is sleeping is 12 watts.  The checkpoint duration is set to two minutes, and the MPI waits are configured as active waits, except otherwise indicated. The scenarios present four processes (or nodes) and the failing node is the one that hosts process 0.

\begin{table}
	\centering
	\caption{Power and slowdown at different clock frequencies}
	\resizebox{\columnwidth}{!}{%
		\begin{tabular}{ |c|c|c|c|c|  }
			\hline
			&  \multicolumn{2}{|c|}{Application} &  \multicolumn{2}{|c|}{Checkpoint}  \\
			\hline
			Frequency (GHz) & Average Power (W) & $\beta$ & Average Power (W) & $\gamma$ \\
			\hline
			2.8 & 166 & 1   & 150 & 1    \\
			2.1 & 148 & 1.2 & 142 &  1.1 \\
			1.7 & 139 & 1.5 & 131 &  1.2 \\
			1.2 & 126 & 2.1 & 125 &  1.4 \\
			\hline
		\end{tabular}
	}
	\label{tab:conf1}
\end{table}

\subsection{Experiments and results analysis}

Different simulated scenarios are discussed in this subsection. For each scenario, the following data are shown: 

\begin{itemize}
	\item A table with the particular configuration data of the experiment, as in Table~\ref{tab:shortLongReexecutionTimeParametros}.
		
	\item A table with the actions selected and the energy savings estimations (as in Table~\ref{tab:shortLongReexecutionTime}): In this table, column $N$ indicates the node number, the $Action$ column indicates the strategy applied, column $T$ indicates the phase duration, and column $TT$ is the total duration. The last columns show the savings in joules, joules per second, and percent ($Save (J)$, $Save Rate (J/s)$, and $Save(\%)$ respectively). The percent column is calculated by dividing the estimations of the joules saved by the joules used without intervention (ie without applying any action). This metric is useful in an intra-
	scenario analyses. To compare different scenarios, J/s is preferred.

	\item The trace graphic: These graphics show the states of each process, the communications between processes, and the time at which any of the corresponding strategies are applied to them. Fig. \ref{fig:estados} indicates the states, communication lines and flags in trace graphics. The red blocks indicate communication blocking or wait (sometimes imperceptible due to the zoom applied to the trace). The flags indicate the beginning and the end of the strategy application in that node. If the strategy applied is to sleep the node, the wait is indicated in gray. Fig. \ref{fig:explicaComm} shows a communication operation between two processes, which is indicated by a yellow line that joins both processes. The inclination of the line (when appreciated) allows us to see who is the sender and who is the receiver. In blocking communications, Fig. \ref{fig:explicaComm} (a), the line goes from the beginning of the send to the end of the receive. In non-blocking communications, Fig. \ref{fig:explicaComm} (b), the line goes from the beginning of the send to the end of the receiver's wait operation (MPI\_Wait). In Fig. \ref{fig:explicaComm} (b), the wait in the receiver is short (the red block is thin) because the wait operation was successful; otherwise, it would be longest.

\begin{figure*}[h]
	\centering
	\includegraphics[width=13cm]{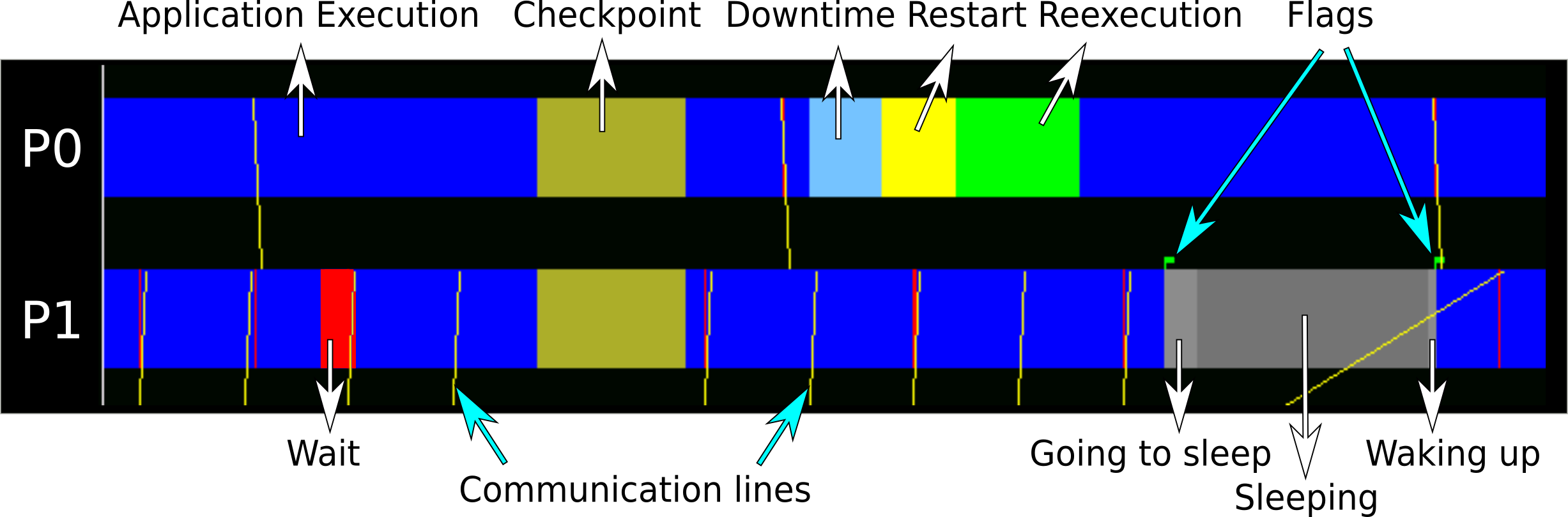}
	\caption{States, communication lines and flags in trace graphics}
	\label{fig:estados}
\end{figure*}

\begin{figure*}[t]
\subfloat[Blocking communication.]{\includegraphics[width=0.45\textwidth,height=0.2\textheight]{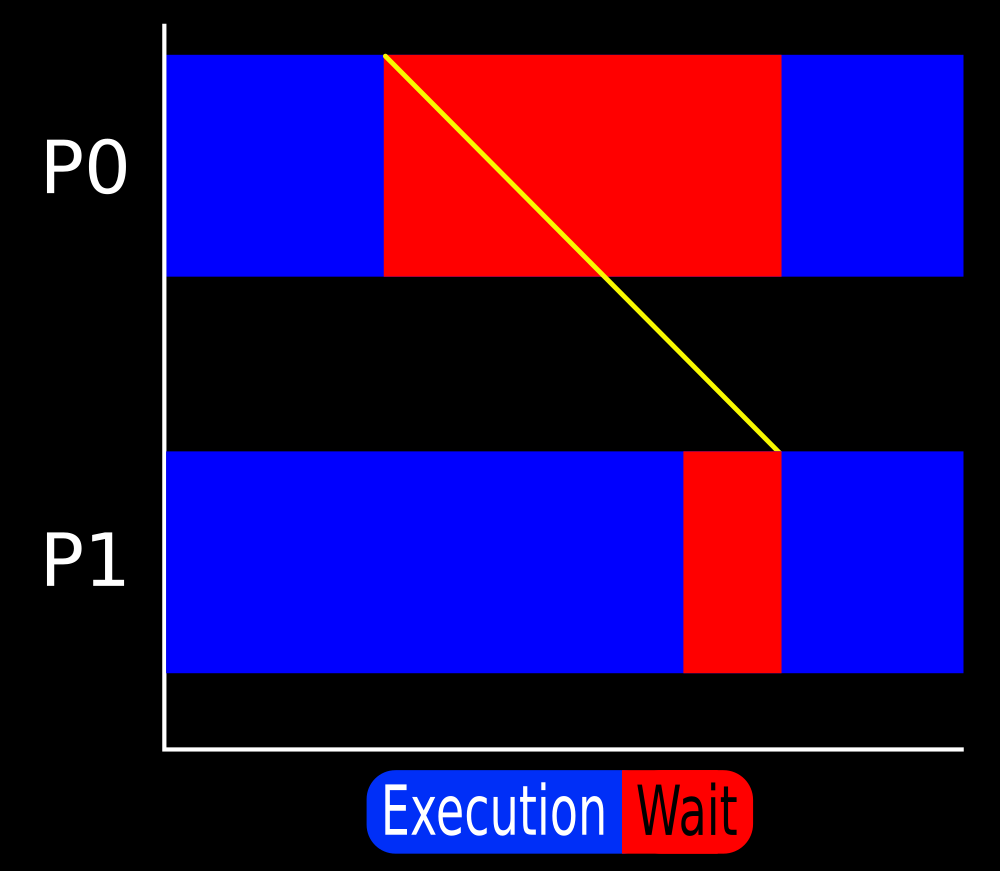}}
\hfill
\subfloat[Non-blocking communication.]{\includegraphics[width=0.45\textwidth,height=0.2\textheight]{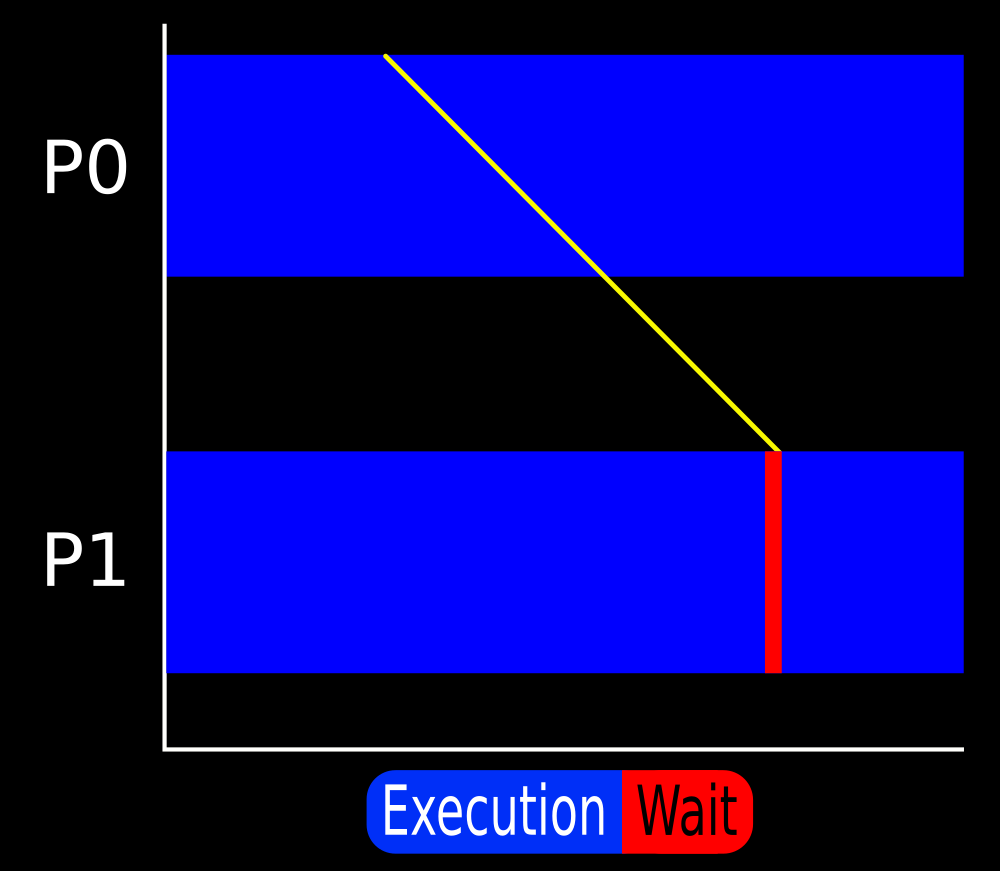}}
	\caption{Communications in trace graphics}
	\label{fig:explicaComm}
\end{figure*}

\end{itemize}

\subsubsection{\textbf{Scenario 1: Short vs. long re-execution time}}
\begin{figure*}[t]
\subfloat[Short re-execution time.]{\includegraphics[width=\textwidth,height=0.29\textheight]{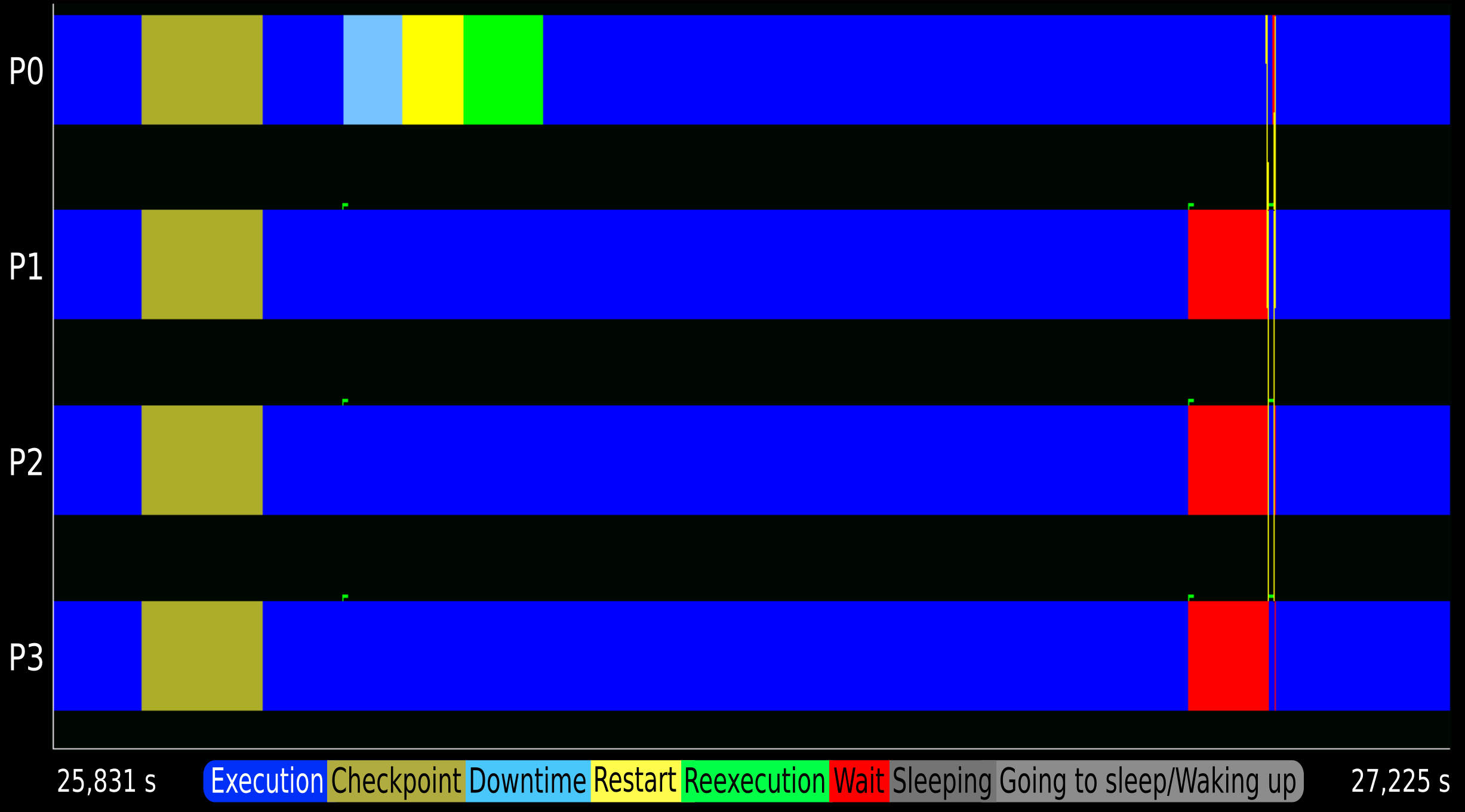}}\\
\subfloat[Long re-execution time.]{\includegraphics[width=\textwidth,height=0.29\textheight]{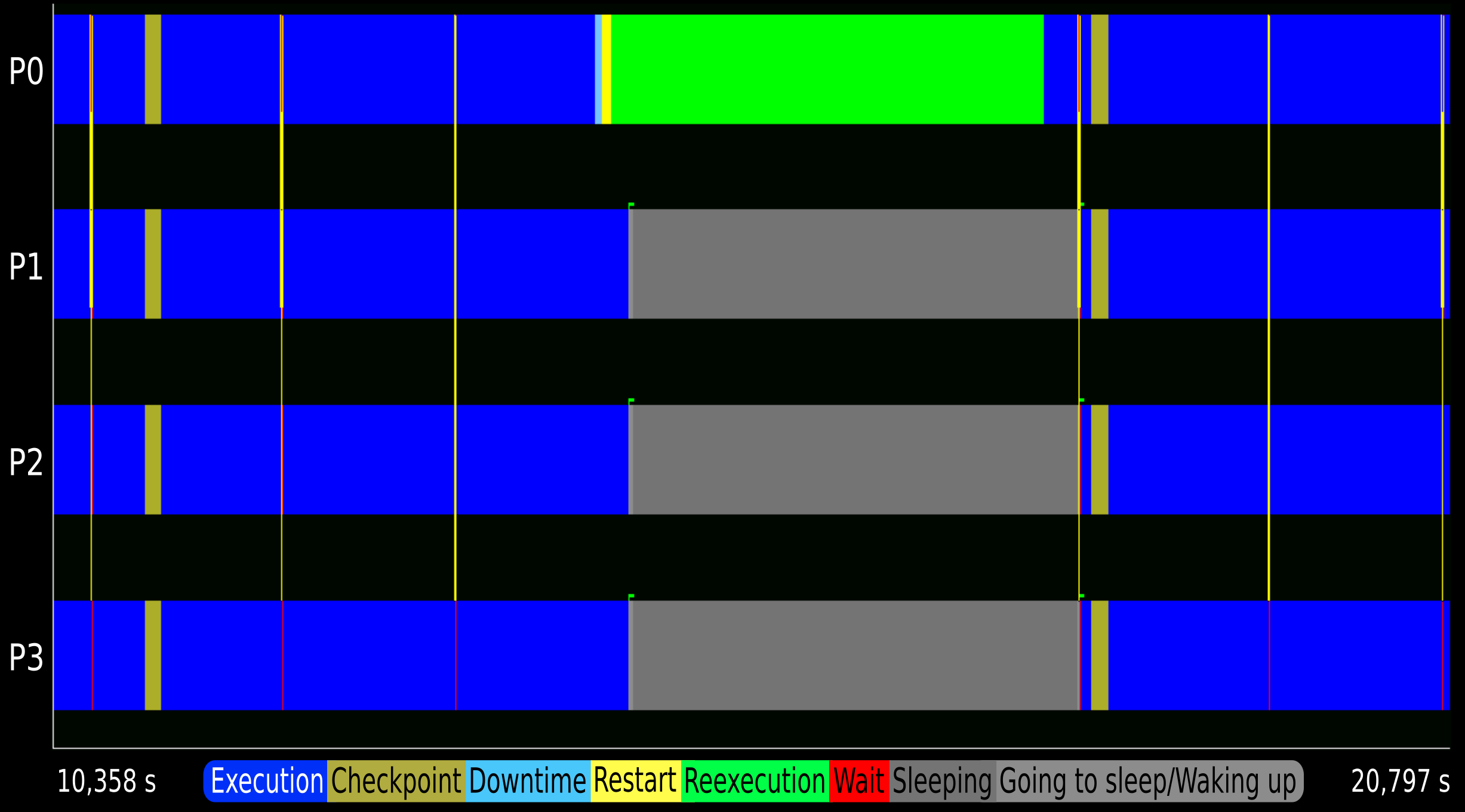}}
	\caption{Scenario 1}
	\label{scenario1}
\end{figure*}

In this scenario, we analyze a case with a long re-execution time compared to one where the re-execution time is short. Table~\ref{tab:shortLongReexecutionTimeParametros} shows the simulation settings. In the short re-execution time case (Fig.~\ref{scenario1}a), the moment of the failure is configured to occur immediately after a checkpoint. The frequency is changed to 2.1 GHz during the computation phase in all processes because lower frequencies increase the synchronization time with the recovering process, and strategies that affect the total execution time are not applied. As this scenario is configured with active waits, and the wait duration is long enough to justify a frequency change (but not to send to sleep), it is changed to the minimum frequency to save energy. With these actions, the three nodes would achieve an energy saving of 14\% in a time interval of 13.4 minutes

The long re-execution time case can be seen in Fig.~\ref{scenario1}b. In this case, the time of failure has been set far from the checkpoint time. As the waiting phases of the three surviving processes are very long, the three nodes are sent to sleep at this phase. Since the nodes will sleep in the waiting phase, it is convenient to arrive at this phase as soon as possible, to obtain greater energy savings. In this scenario, the clock frequencies different from the maximum one implies a longer compute phase, and therefore a shorter waiting phase. This leads to fewer energy savings and this is the reason why the clock frequency is not changed in the compute phase. With these actions, the intervened nodes would be able to consume almost 86\% less energy during the intervention interval, which is around 60 minutes. This scenario achieves better results than the previous one due to its long waits where the nodes go to sleep. Figs.~\ref{scenario1}a and~\ref{scenario1}b are at different scales to better appreciate each case. Table~\ref{tab:shortLongReexecutionTime} shows the results obtained.

\begin{table}[]
	\centering
	\caption{Simulation parameters for Scenario 1: Short vs. long re-execution time}
	\resizebox{\columnwidth}{!}{%
		\begin{tabular}{ |l|l| }
			\hline
			Communication interval & 21.6 min. \\
			\hline
			Communication pattern & Process P0 sends and receives messages from processes P1, P2, and P3. \\
			\hline
			MPI Waits & Active     \\
			\hline
			MPI operations & Blocking   \\
			\hline		
		\end{tabular}
	}
	\label{tab:shortLongReexecutionTimeParametros}
\end{table}

\begin{table}[]
	\centering
	\caption{Selected actions and energy savings for scenario 1: Short vs. long re-execution time}
	\resizebox{\columnwidth}{!}{%
		\begin{tabular}{ |c|c|r|c|r|r|r|r|r| }
			\hline
			&  \multicolumn{2}{|c|}{Compute phase} &  \multicolumn{2}{|c|}{Wait phase} & \multicolumn{4}{|c|}{}  \\
			\hline
			N & Action & T (m) & Action & T (m) & TT (m) & Save (J)   & Save Rate(J/s) & Save (\%) \\
			\hline
			\multicolumn{9}{|c|}{Short re-execution time} \\
			\hline
			1 & 2.1 GHz    & 12.07  & 1.2 GHz 	& 1.32  & 13.39  & 18,704.5		& 23.28    & 14.03     \\
			2 & 2.1 GHz    & 12.07  & 1.2 GHz  	& 1.32  & 13.39  & 18,705.56  	& 23.28    & 14.02     \\
			3 & 2.1 GHz    & 12.07  & 1.2 GHz  	& 1.32  & 13.39  & 18,706.06  	& 23.28    & 14.02     \\
			\hline
			\multicolumn{9}{|c|}{Long re-execution time} \\
			\hline
			1 & No action  & 4.37 & sleep  & 56.00 & 60.37  & 516,084.73 & 153.59     & 85.82     \\
			2 & No action  & 4.37 & sleep  & 56.00 & 60.38  & 516,085.34 & 153.59     & 85.82     \\
			3 & No action  & 4.38 & sleep  & 56.00 & 60.38  & 516,084.69 & 153.59     & 85.82     \\
			\hline
		\end{tabular}
	}
	\label{tab:shortLongReexecutionTime}
\end{table}

% 22E_reejecCorta/New_window_#8@traza_22E.png

\subsubsection{\textbf{Scenario 2: Blocking vs. non-blocking MPI operations}}

\begin{figure*}[t]
\subfloat[Blocking operations.]{\includegraphics[width=\textwidth,height=0.29\textheight]{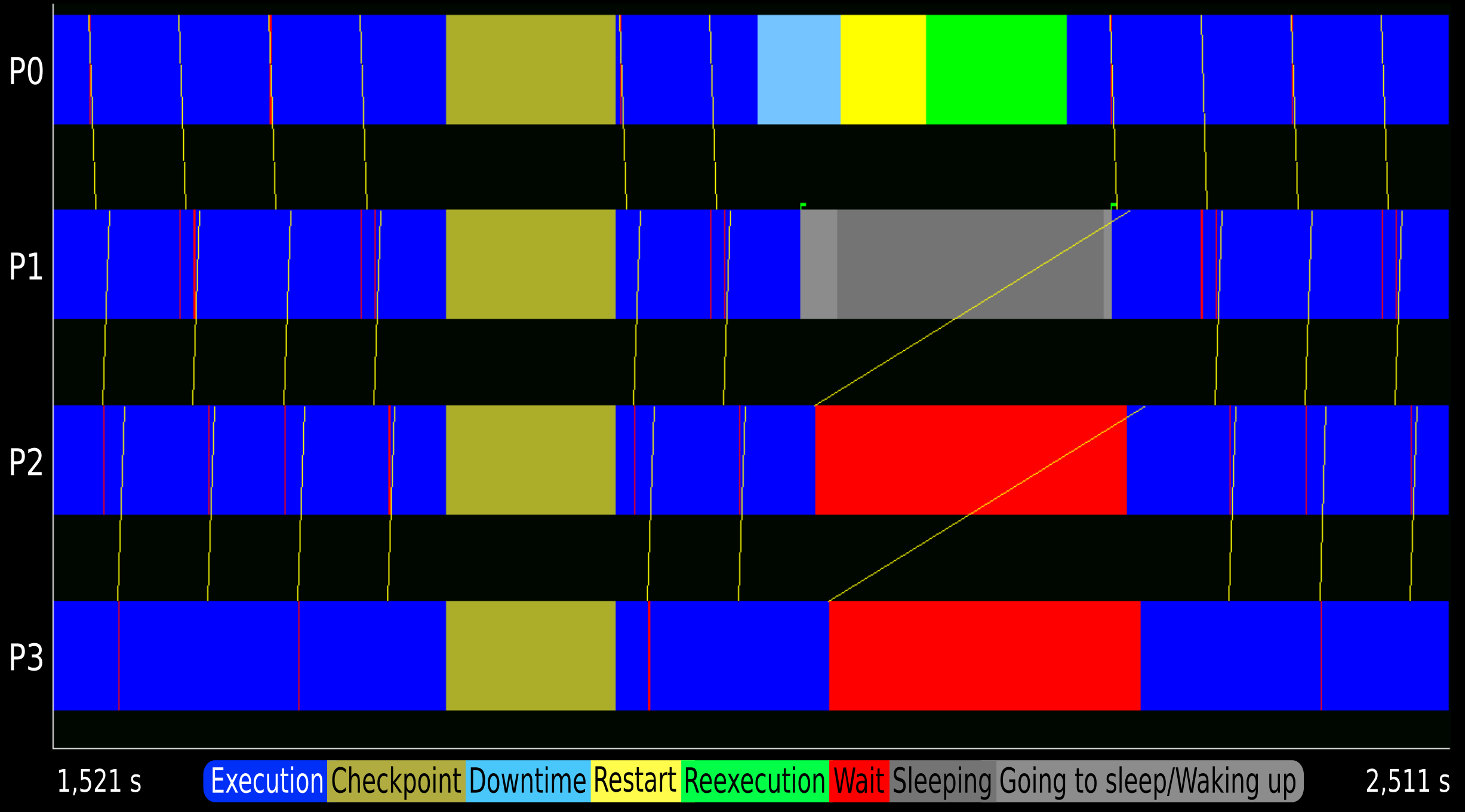}}\\
\subfloat[ Non-blocking operations.]{\includegraphics[width=\textwidth,height=0.29\textheight]{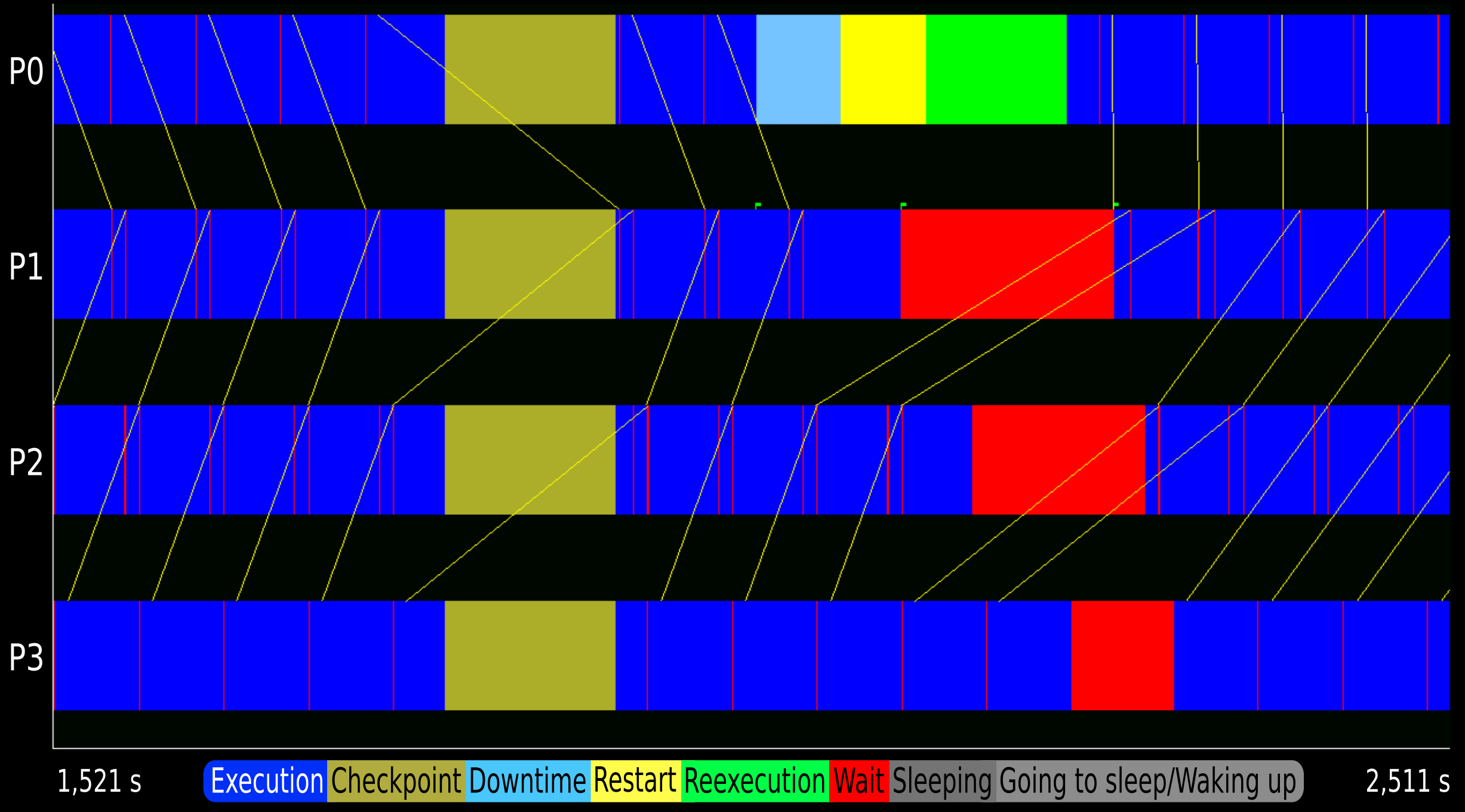}}
	\caption{Scenario 2}
	\label{scenario2}
\end{figure*}

In this scenario, we compare the use of blocking and non-blocking operations. Table~\ref{tab:BloqueantesNoBloqueantesParametros} shows the configuration of the simulation. Fig.~\ref{scenario2}a shows the use of blocking operations. In this figure, after the failure, process P0 cannot send data to process P1. Without this data, process P1 cannot continue its computation and gets blocked. At this moment process P1 goes to sleep, achieving a saving of 72\% in almost 5 minutes.

Fig.~\ref{scenario2}b shows the use of non-blocking operations. In this case, after the failure, process P1 can complete its communication with Process P0 (the $wait$ is successful) because it had been done just before the failure. Then, it continues computing and blocks at the next $wait$. This results in a longer computational phase when compared to the previous case. As can be seen in Figure~\ref{scenario2}b, process P1 changed the frequency in its computation and waiting phase (indicated by the green flags in the figure), achieving a saving of almost 30\% in an interval time similar to the previous case (4 minutes and a half).

This difference in power savings is because the node sleeps in the first case (use of blocking operations), while it changes clock frequency in the second case (use of non-blocking operations). Table~\ref{tab:BloqueantesNoBloqueantes} shows the results obtained.

\begin{table}[t]
	\centering
	\caption{Simulation parameters for scenario 2: Blocking vs. non-blocking MPI operations }
	\resizebox{\columnwidth}{!}{%
		\begin{tabular}{ |l|l| }
			\hline
			Communication interval & 5 min. \\
			\hline
			Communication pattern & Processes P0 and P2 send messages to process 1. Process P3 sends messages to process P2 \\
			\hline
			MPI Waits & Active     \\
			\hline
			MPI operations & Blocking   \\
			\hline		
		\end{tabular}
	}
	\label{tab:BloqueantesNoBloqueantesParametros}
\end{table}

\begin{table}[t]
	\centering
	\caption{Selected actions and energy savings for scenario 2: Blocking vs. non-blocking MPI operations}
	\resizebox{\columnwidth}{!}{%
		\begin{tabular}{ |c|c|r|c|r|r|r|r|r| }
			\hline
			&  \multicolumn{2}{|c|}{Compute phase} &  \multicolumn{2}{|c|}{Wait phase} & \multicolumn{4}{|c|}{}  \\
			\hline
						N & Action & T (m) & Action & T (m) & TT (m) & Save (J)   & Save Rate(J/s) & Save (\%) \\
			\hline
			\multicolumn{9}{|c|}{ Blocking operations} \\
			\hline

				1 & No action   & 0.86  & sleep  & 3.67 & 4.53  & 32,502.30 & 147.77 & 72.06     \\
							\hline
			\multicolumn{9}{|c|}{ Non-blocking operations} \\
			\hline
			1 & 2.1 GHz     & 2.07  & 1.2 GHz  & 2.15 & 4.22  & 11,468.73 & 45.30 & 27.29     \\
			\hline
		\end{tabular}
	}
	\label{tab:BloqueantesNoBloqueantes}
\end{table}

% 4_y_medio_E_sin_cascada_sync_4p_OK/New_window_#11@traza_4_y_medioE_1_sync_correjido.png
%4_y_medio_E_sin_cascada_async_4p_OK/New_window_#12@traza_4_y_medioE_1_async_correjido.png

\subsubsection{\textbf{Scenario 3: Active waits vs. idle waits}}
\begin{figure*}[t]
\subfloat[Active waits.]{\includegraphics[width=\textwidth,height=0.29\textheight]{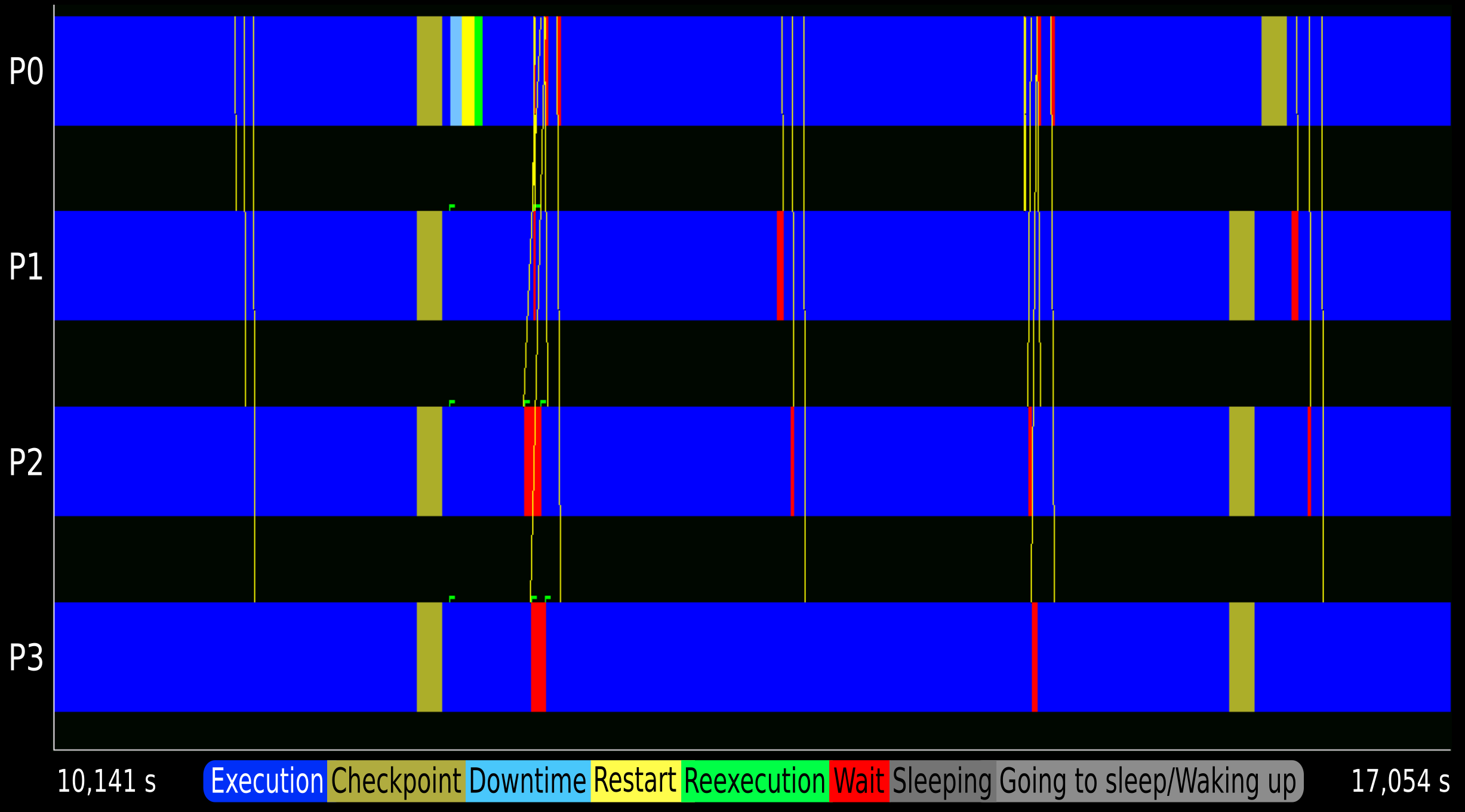}}\\
\subfloat[Idle waits.]{\includegraphics[width=\textwidth,height=0.29\textheight]{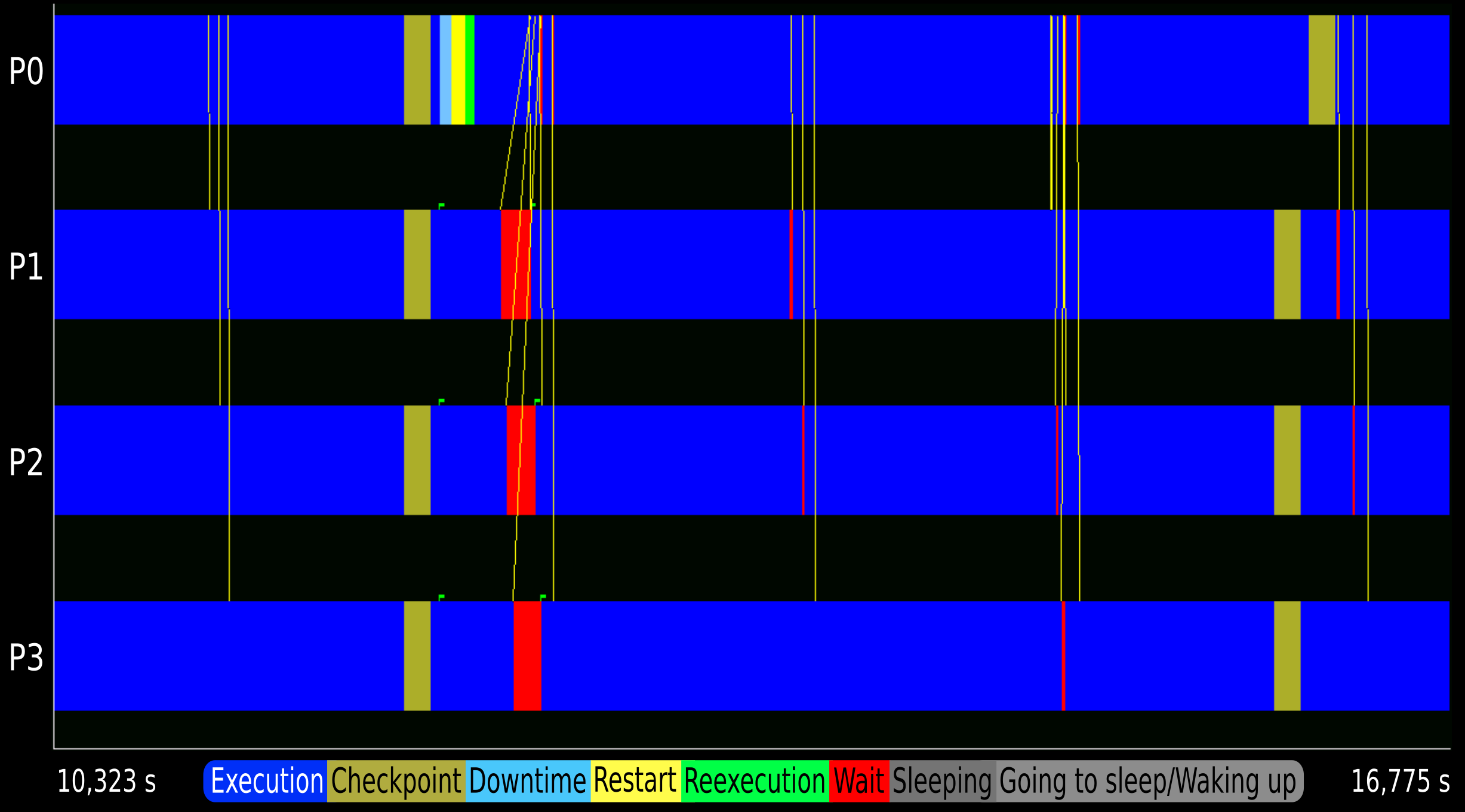}}
	\caption{Scenario 3}
	\label{scenario3}
\end{figure*}

In this scenario, we compare the use of active waits against idle waits (see subsection~\ref{sub:esperasMPI}). Table~\ref{tab:EsperaActivaYnoActivaParametros} shows the simulation configuration. In both cases, the actions selected for the computation phase are to change the frequency. In the active wait case (Fig.~\ref{scenario3}a), the action selected for the waiting phase is to change the frequency to the minimum one, indicated by the green flag at the beginning of the phase. In the idle wait case (Fig.~\ref{scenario3}b), there is no action in the waiting phase. 

In the first case (active waits), the energy savings are 36\%, against the 0.09\% for the second case (idle waits), in the same interval of time. This shows the positive impact of the application of the strategy on active waits. Even when the scenario presents short waiting times, if the system is configured with active waiting, the use of the strategies achieves considerable energy savings. Table~\ref{tab:EsperaActivaYnoActiva} shows the results obtained.
 
\begin{table}[t]
	\centering
	\caption{Simulation parameters for scenario 3: Active waits vs. idle waits }
	\resizebox{\columnwidth}{!}{%
		\begin{tabular}{ |l|l| }
			\hline
			Communication interval & 60 sec. \\
			\hline
			Communication pattern & Process P0 sends and receives messages from processes P1, P2, and P3.   \\
			\hline
			MPI Waits & Active and non-active    \\
			\hline
			MPI operations & Blocking   \\
			\hline
		\end{tabular}
	}
	\label{tab:EsperaActivaYnoActivaParametros}
\end{table}

\begin{table}[t]
	\centering
	\caption{Selected actions and energy savings for scenario 3: Active waits vs. idle waits}
	\resizebox{\columnwidth}{!}{%
		\begin{tabular}{ |c|c|r|c|r|r|r|r|r| }
			\hline
			&  \multicolumn{2}{|c|}{Compute phase} &  \multicolumn{2}{|c|}{Wait phase} & \multicolumn{4}{|c|}{}  \\
			\hline
			N & Action & T (m) & Action & T (m) & TT (m) & Save (J)   & Save Rate(J/s) & Save (\%) \\
			\hline
			\multicolumn{9}{|c|}{ Active waits} \\
			\hline
			1 & 2.1 GHz   & 0.64  & 1.2 GHz   & 2.56 & 3.20  & 11,673.84 & 60.78 & 36.61     \\
			\hline
			2 & 2.1 GHz   & 0.64  & 1.2 GHz   & 2.56 & 3.20  & 11,674.89 & 60.75 & 36.60     \\
			\hline
			3 & 2.1 GHz   & 0.65  & 1.2 GHz   & 2.56 & 3.21  & 11,675.40 & 60.72 & 36.58     \\
			\hline
			\multicolumn{9}{|c|}{ Non-active waits} \\
			\hline
			1 & 2.1 GHz   & 0.64  & No action  & 2.56 & 3.20  & 12.83 & 0.33 & 0.09     \\
			\hline
			2 & 2.1 GHz   & 0.64  & No action  & 2.56 & 3.20  & 12.87 & 0.33 & 0.09     \\
			\hline
			3 & 2.1 GHz   & 0.65  & No action  & 2.56 & 3.21  & 12.92 & 0.33 & 0.09     \\
			\hline
		\end{tabular}
	}
	\label{tab:EsperaActivaYnoActiva}
\end{table}

\subsubsection{\textbf{Scenario 4: With and without system buffers}}

In this scenario, we analyze the effects of using or not using system buffers during communication operations (see subsection~\ref{sub:operationsMPI}). Tabla~\ref{tab:ConysinbufferParametros} shows the simulation configuration. As indicated in this table, processes 1, 2, and 3 send data to process 0.

\begin{figure*}[t]
\subfloat[Without system buffers.]{\includegraphics[width=\textwidth,height=0.29\textheight]{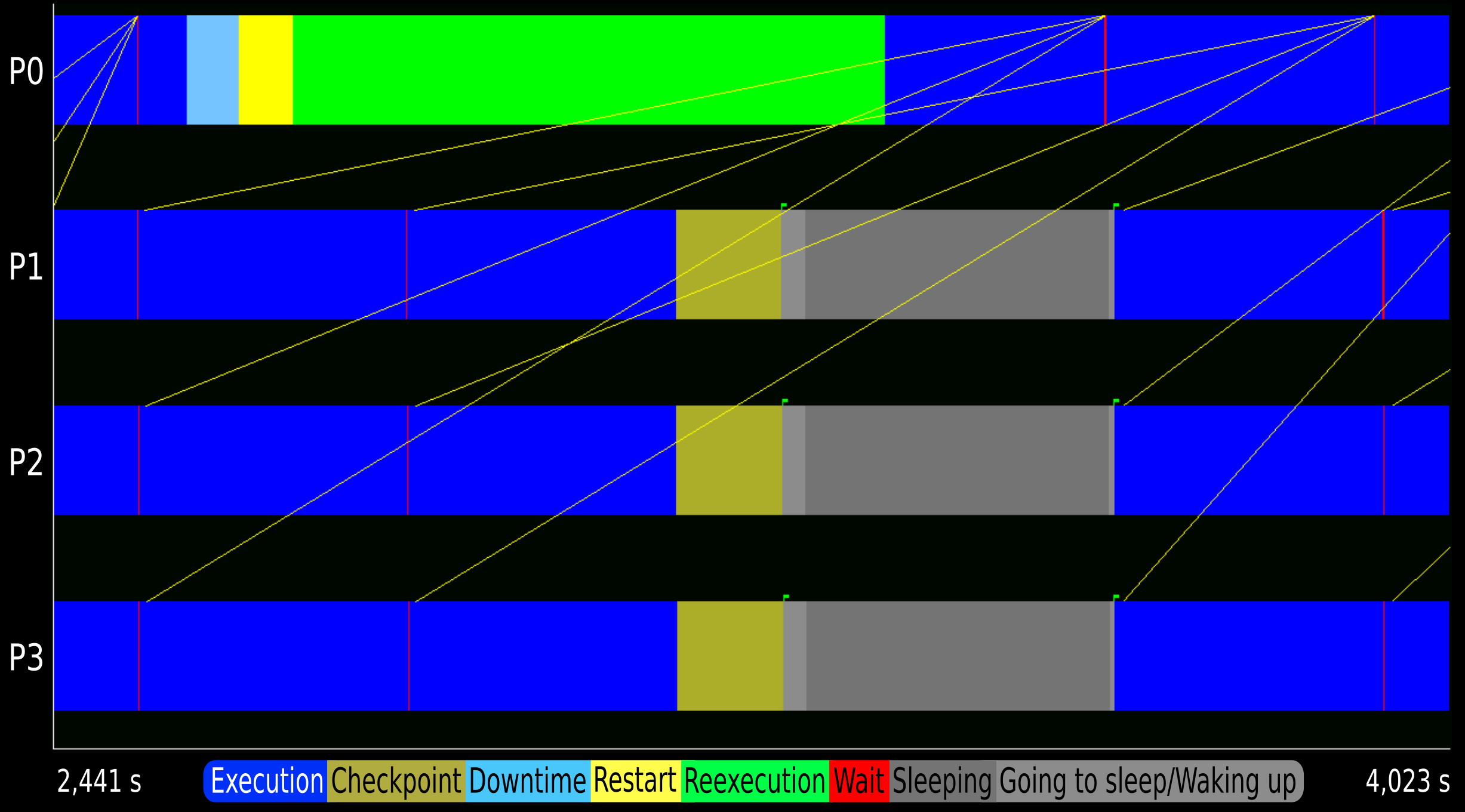}}\\
\subfloat[With system buffers.]{\includegraphics[width=\textwidth,height=0.29\textheight]{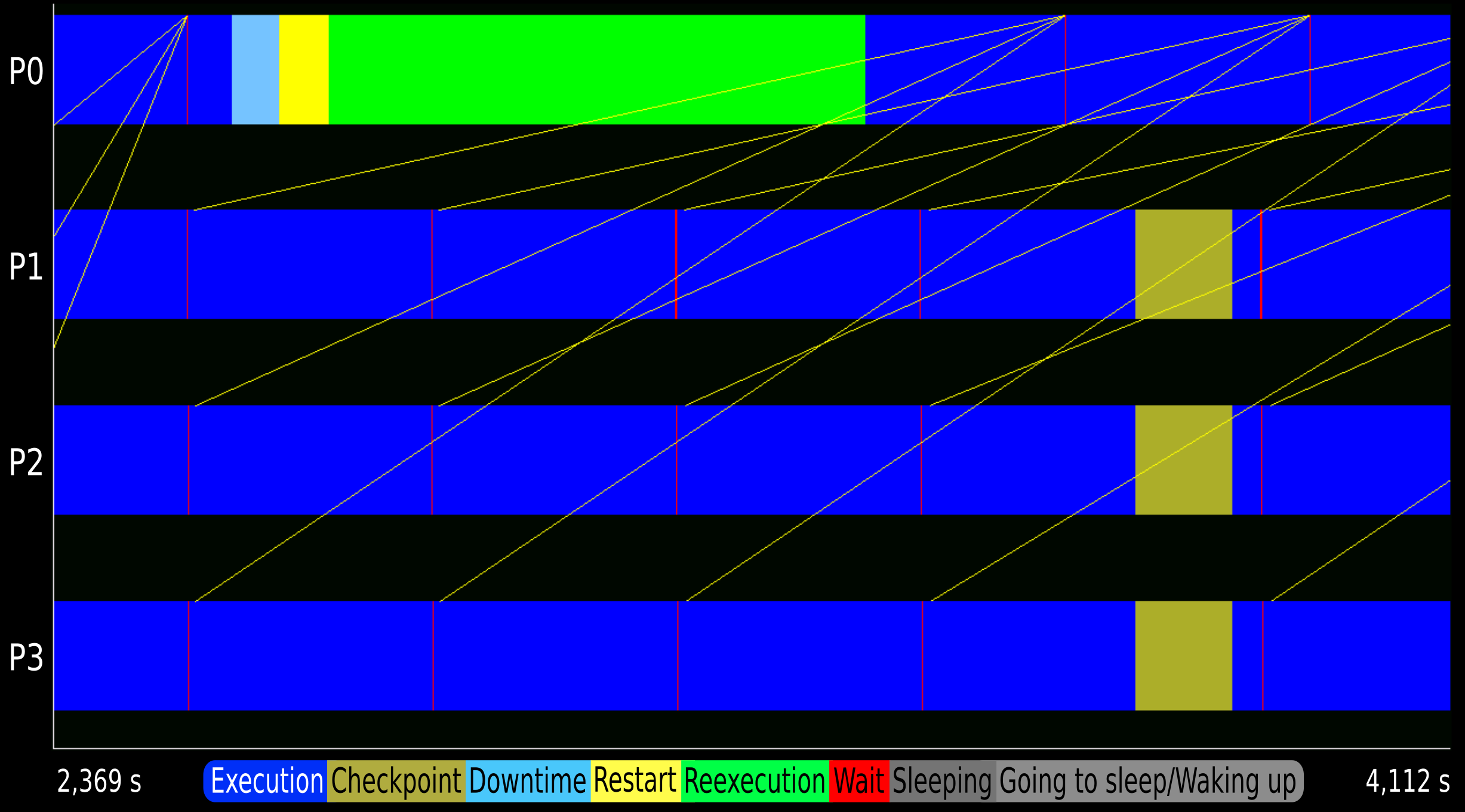}}
	\caption{Scenario 4}
	\label{scenario4}
\end{figure*}

Fig.~\ref{scenario4}a shows the case of non-blocking operations without buffering. As it can be seen, after the failure, the next $wait$ operation of the non-failing processes is successful because the communication had taken place just before the failure. On the next $wait$ operation, all three processes get blocked because the receiving process (Process P0) is still re-executing. The selected strategy for the three processes is to keep the frequency in the compute phase and sleep in the waiting phase. The observed energy saving is approximately 30\% in an interval of 17 minutes.

The system buffers store the sent messages so that several $send$ operations (in the case of blocking operations) or several $wait$ operations (in the case of non-blocking operations) can be issued successively without blocking. Fig.~\ref{scenario4}b shows a simulation where system buffering is used. As it can be seen, the chosen communication pattern (living processes send data to the failing process) avoids the blocking of live processes, which are not affected by the failure, and therefore do not receive the application of any of the strategies.

Although the proposed experiment is quite atypical, it serves to magnify and observe the impact that the presence or absence of system buffering has on communications operations when an application becomes desynchronized due to the failure of a node. This is the reason
why the simulator is able to enable or disable the use of the system buffering in communication operations. In a typical message-passing application, the processes usually alternate between send and receive operations, so that if the $send$ is not blocked (due to the use of the system buffer), the $receive$ will be blocked, and the processes can receive the application of the strategies. Beyond the savings achieved in one case or another, the use of the buffer allows the processes to advance in doing useful work, which has a positive impact on energy efficiency. Table~\ref{tab:ConYsinBuffer} shows the results obtained.

\begin{table}[t]
	\centering
	\caption{Simulation parameters for scenario 4: With and without system buffers }
	\resizebox{\columnwidth}{!}{%
		\begin{tabular}{ |l|l| }
			\hline
			Communication interval & 5 min. \\
			\hline
			Communication pattern & Processes P1, P2, and P3 send messages to process P0. \\
			\hline
			MPI Waits & Active     \\
			\hline
			MPI operations & Non-blocking   \\
			\hline		
		\end{tabular}
	}
	\label{tab:ConysinbufferParametros}
\end{table}

\begin{table}[t]
	\centering
	\caption{Selected actions and energy savings for scenario 4: With and without system buffers }
	\resizebox{\columnwidth}{!}{%
		\begin{tabular}{ |c|c|r|c|r|r|r|r|r| }
			\hline
			&  \multicolumn{2}{|c|}{Compute phase} &  \multicolumn{2}{|c|}{Wait phase} & \multicolumn{4}{|c|}{}  \\
			\hline
			N & Action & T (m) & Action & T (m) & TT (m) & Save (J)   & Save Rate (J/s) & Save (\%) \\
			\hline
			\multicolumn{9}{|c|}{Without system buffer} \\
			\hline
			1 & No action   & 11.25  & sleep  & 6.27 & 17.51  & 56549.4 & 150.36 & 32.78     \\
			\hline
			2 & No action     & 11.26  & sleep  & 6.26 & 17.52  & 56426.2 & 150.35 & 32.7     \\
			\hline
			3 & No action     & 11.28  & sleep  & 6.24 & 17.52  & 56303 & 150.34 & 32.62     \\
			\hline
			
		\end{tabular}
	}
	\label{tab:ConYsinBuffer}
\end{table}

%14E_both_async_sinBuffer_OK
%15E_both_async_conBuffer_OK

\subsubsection{\textbf{Scenario 5: With and without analysis of cascade-blocked processes}}

\begin{figure*}[t]
\subfloat[Without analysis of cascade-blocked processes]{\includegraphics[width=\textwidth,height=0.29\textheight]{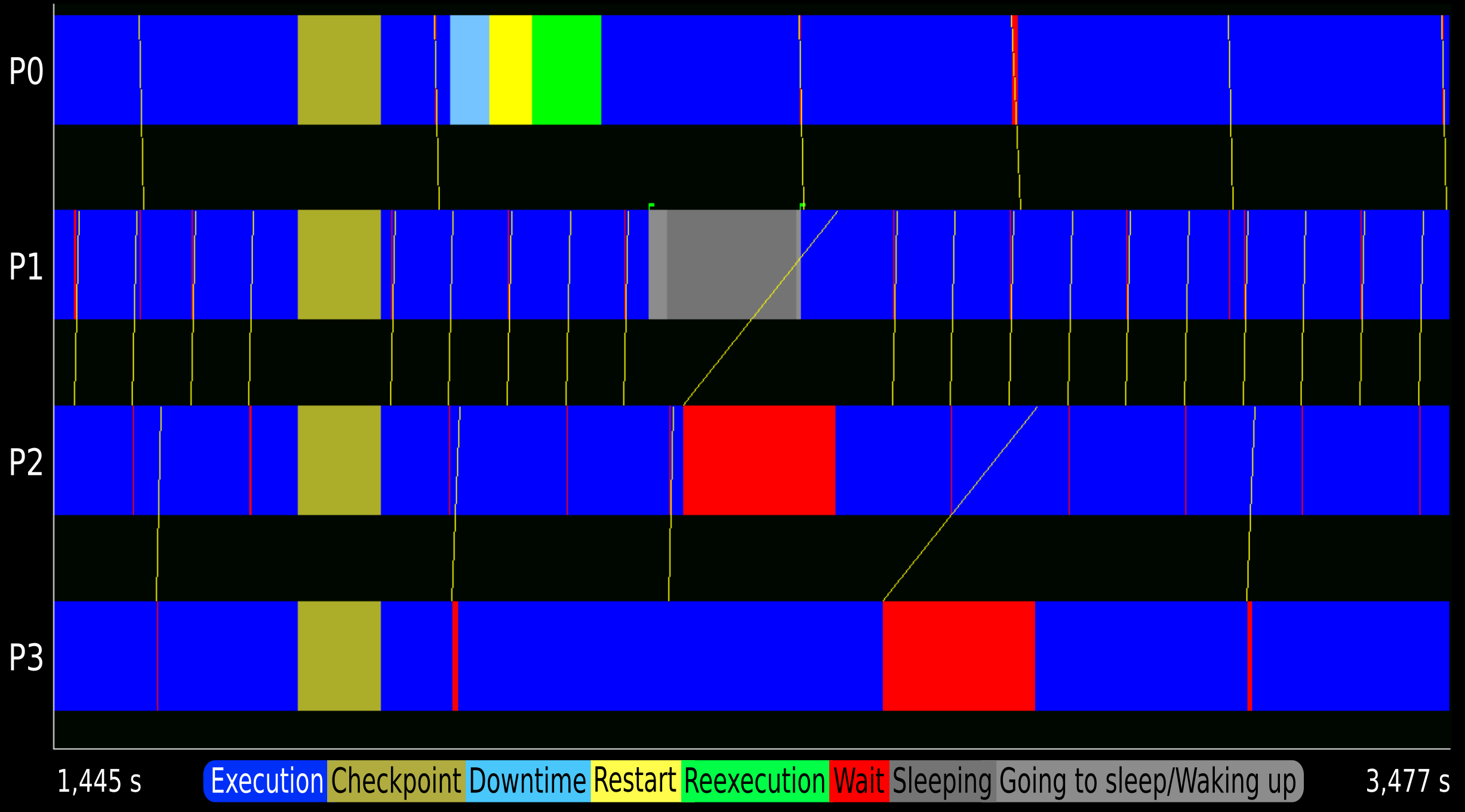}}\\
\subfloat[With analysis of cascade-blocked processes]{\includegraphics[width=\textwidth,height=0.29\textheight]{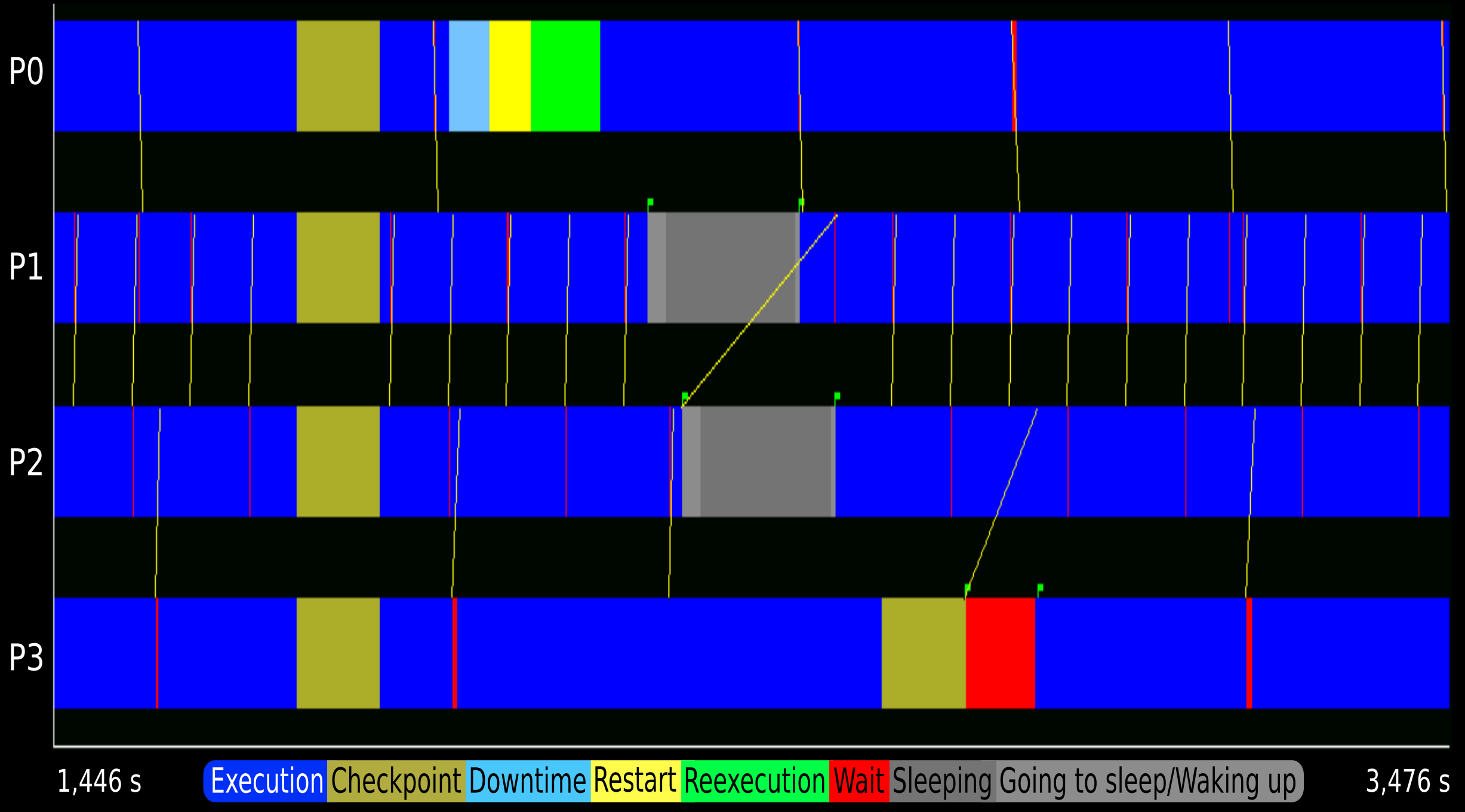}}
	\caption{Scenario 5}
	\label{scenario5}
\end{figure*}

This scenario analyzes the effect of incorporating cascade-blocked processes (see section~\ref{sec:politicas}) into the processes that receive the application of the strategies. Table~\ref{tab:CascadaParametros} shows the configuration of the simulation. Fig.~\ref{scenario5}a shows a simulation that does not analyze cascade blocked processes. As it can be seen, process P1 blocks with process P0 (failing process) and receives the application of a strategy (sleep). However, processes P2 and P3 get also blocked because of the failure, but in an indirect manner. By including cascade-blocked processes in the analysis, we can note that process P2 communicates with process P1 four times before it get blocked, so we need to raise the depth level to 5 (see section~\ref{sec:politicas} for a definition of $depth$). Fig.~\ref{scenario5}b shows that the blocking state of process P2 is now detected, and the strategy selected in this case is to send the node to sleep. When this blocking state is included in the analysis, the blocking state of process P3 is also detected, which is cascade-blocked with process P2. The strategy applied to process P3 is to change the frequency during the waiting phase. Although process P3 blocks on the third communication with process P2, and a depth of 3 would have been enough, it was necessary to increment the depth to 5 to include the mentioned blocking state in the analysis (because of its relation with process P2).

By including the cascade-blocked processes in the simulator analysis, we cover from a scenario where only one process takes an action to improve energy efficiency (Fig.~\ref{scenario5}a), to another where the three living processes do the same (Fig.~\ref{scenario5}b). Analyzing the cascade-blocked processes, allow us to increase energy savings from 32,500 J to almost 80,000 J in intervals between 9 and 14 minutes. Table~\ref{tab:Prof1y5} shows the results obtained.

\begin{table}[t]
	\centering
	\caption{Simulation parameters for scenario 5: With and without analysis of cascade-blocked processes}
	\resizebox{\columnwidth}{!}{%
		\begin{tabular}{ |l|l| }
			\hline
			Communication interval & 5 sec. \\
			\hline
			Communication pattern & Processes 0 and 2 send messages to process 1, and process 3 sends messages to process 2.   \\
			\hline
			MPI Waits & Active     \\
			\hline
			MPI operations & Blocking   \\
			\hline		
		\end{tabular}
	}\label{tab:CascadaParametros}
\end{table}

\begin{table}[t]
	\centering
	\caption{Selected actions and energy savings for scenario 5: With and without analysis of cascade-blocked processes}
	\resizebox{\columnwidth}{!}{%
		\begin{tabular}{ |c|c|r|c|r|r|r|r|r| }
			\hline
			&  \multicolumn{2}{|c|}{Compute phase} &  \multicolumn{2}{|c|}{Wait phase} & \multicolumn{4}{|c|}{}  \\
			\hline
			Node & Action & T (m) & Action & T (m) & TT (m) & Save (J) & Save Rate (J/s) & Save (\%) \\
			\hline
			\multicolumn{9}{|c|}{Without analysis of cascading blocked processes} \\
			\hline
			1 & No action     & 4.84 & sleep  & 3.67 & 8.51  & 32,517.7 & 147.77 & 38.36     \\
			\hline		
			\multicolumn{9}{|c|}{With analysis of cascading blocked processes} \\
			\hline
			1 & No action     & 4.51  & sleep  & 4.00 & 8.51  & 35,597.70 & 148.29 & 42.00     \\
			2 & No action     & 4.93  & sleep  & 4.00 & 8.93  & 35,636.20 & 148.30 & 40.03     \\
			3 & No action     & 12.00  & 1.2 GHz  & 02.02 & 14.02  & 8,655.07 & 71.5 & 6.29     \\
			\hline
		\end{tabular}
	}
	\label{tab:Prof1y5}
\end{table}

%18E_cascada_analisis_prof: New_window_#4@traza_18E_E5_Prof1.png
%18E_cascada_analisis_prof: New_window_#4@traza_18E_E5_Prof5.png

\subsubsection{\textbf{Scenario 6: With and without checkpoint anticipations}}

This scenario analyzes the impact of bringing forward checkpoints. Table~\ref{tab:checkAheadParametros} shows the configuration of the simulation. This scenario compares two cases with a long re-execution time. 

Fig. 10 a) and b) show the cases when the
advanced checkpoint anticipations is enabled and
disabled in the simulator, respectively.

\begin{figure*}[t]
\subfloat[With checkpoint anticipations]{\includegraphics[width=\columnwidth,height=0.29\textheight]{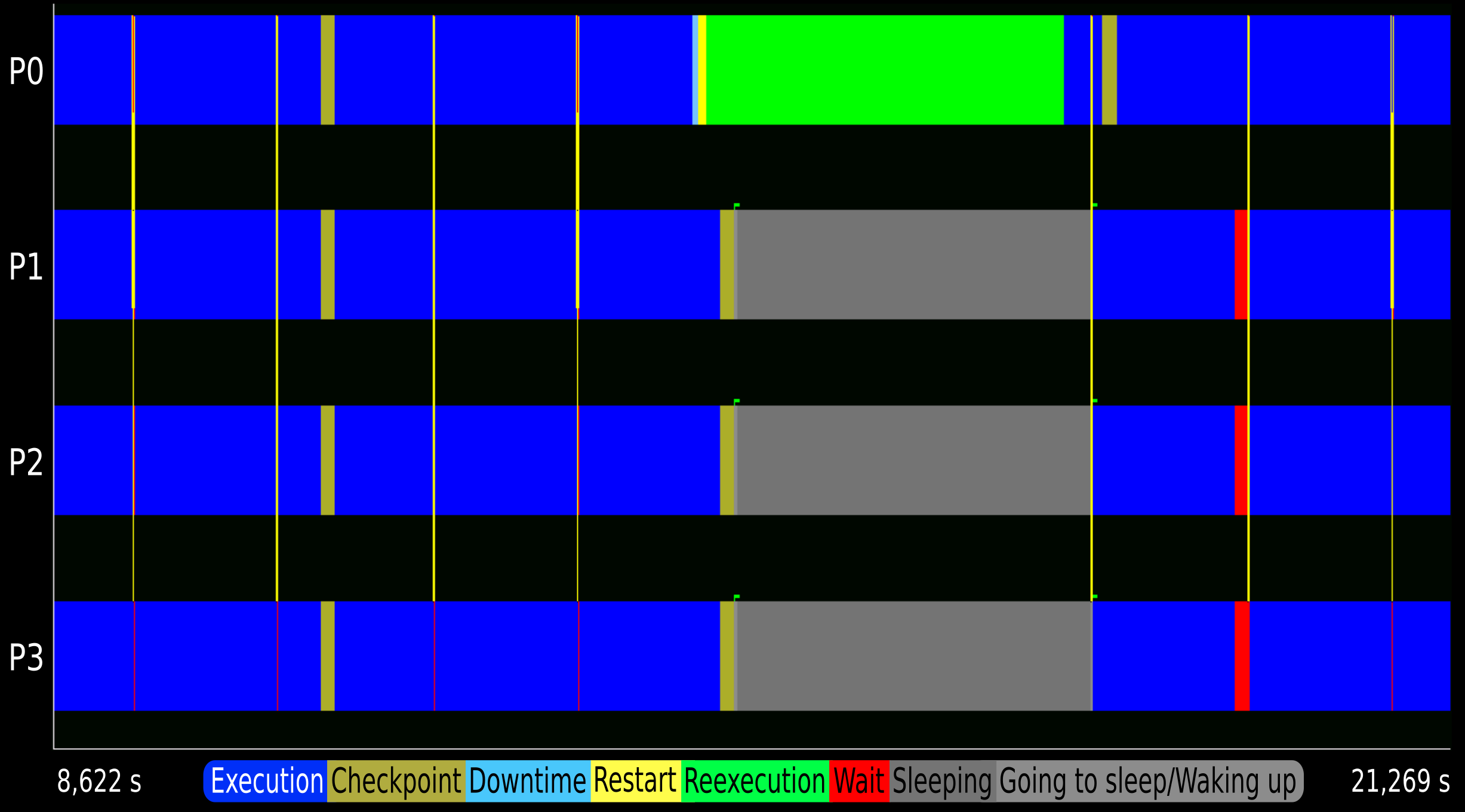}}\\
\subfloat[ Without checkpoint anticipations]{\includegraphics[width=\columnwidth,height=0.29\textheight]{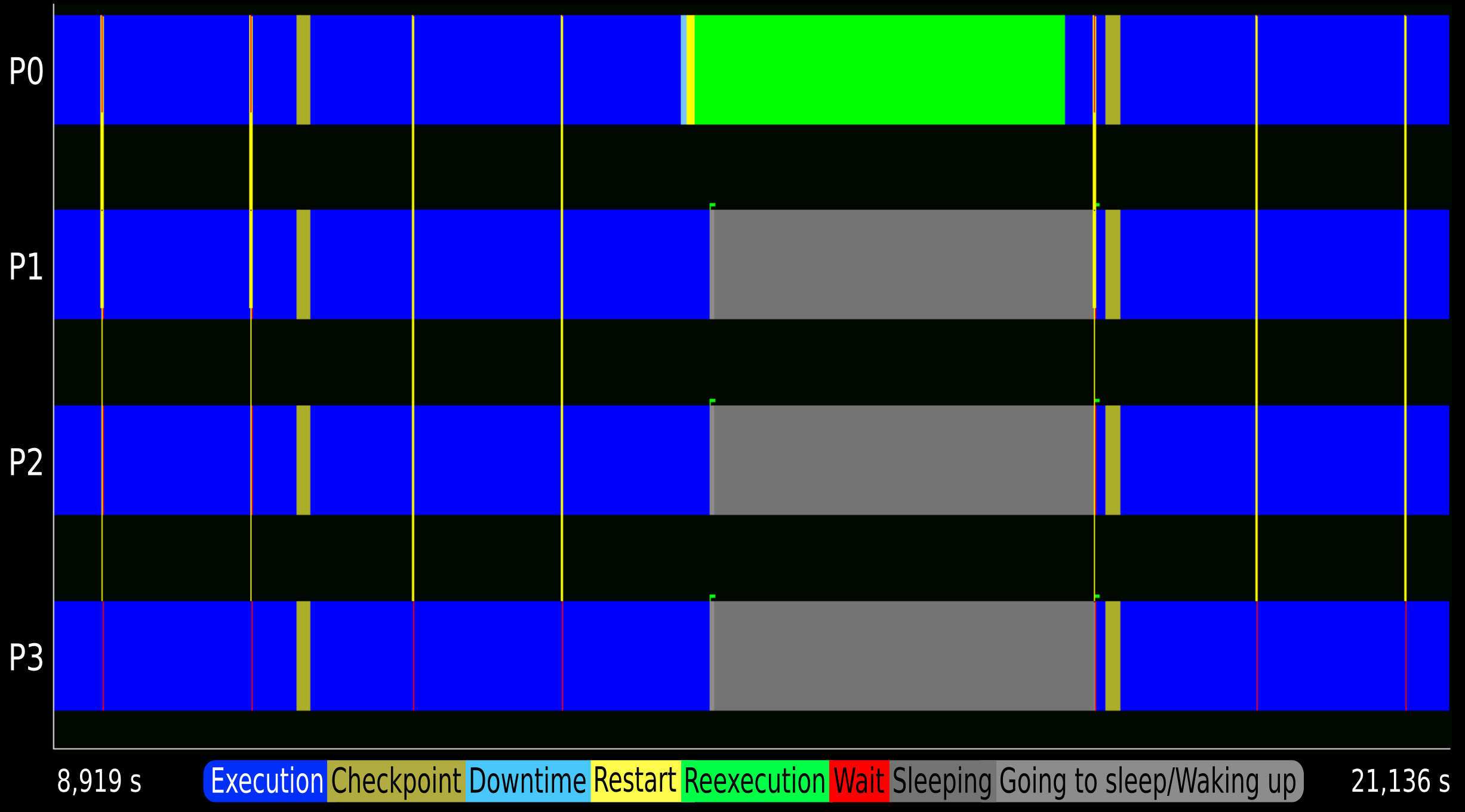}}
	\caption{Scenario 6}
	\label{scenario6}
\end{figure*}

Figs.~\ref{scenario6}a and~\ref{scenario6}b show the cases when checkpoint anticipation is enabled and disabled in the simulator, respectively. The selected strategy in both cases is to maintain the maximum frequency in the compute phase and sleep the nodes in the waiting phase. In the first case, before entering the waiting phase, checkpoints are anticipated. This causes a shorter waiting phase than the second case. The results show slightly higher energy savings in the case where no checkpoint is advanced, (85.8\% versus 83\%). Although there is a slightly improves in the first case (without advanced checkpoints), in the second case (with advanced checkpoints) it avoids stopping the application later when it might be doing useful computation. Table~\ref{tab:checkAdelantado} shows the results obtained.

\begin{table}[t]
	\centering
	\caption{Simulation parameters for scenario 6:  With and without checkpoint anticipations}
	\resizebox{\columnwidth}{!}{%
		\begin{tabular}{ |l|l| }
			\hline
			Communication interval & 4 sec. \\
			\hline
			Communication pattern & Process P0 sends and receives messages from processes P1, P2, and P3.   \\
			\hline
			MPI Waits & Active     \\
			\hline
			MPI operations & Blocking   \\
			\hline		
		\end{tabular}
	}
	\label{tab:checkAheadParametros}
\end{table}

\begin{table}[t]
	\centering
	\caption{Selected actions and energy savings for scenario 6:  With and without checkpoint anticipations}
	\resizebox{\columnwidth}{!}{%
		\begin{tabular}{ |c|c|r|c|r|r|r|r|r| }
			\hline
			&  \multicolumn{2}{|c|}{Compute phase} &  \multicolumn{2}{|c|}{Wait phase} & \multicolumn{4}{|c|}{}  \\
			\hline
			Node & Action & T (m) & Action & T (m) & TT (m) & Save (J) & Save Rate (J/s) & Save (\%) \\
			\hline
			\multicolumn{9}{|c|}{With moving ahead checkpoints} \\
			\hline
			1 & No action   & 6.37  & sleep  & 54.00 & 60.37  & 497.604.73 & 153.58     & 83.02     \\
			2 & No action   & 6.37  & sleep  & 54.00 & 60.38  & 497,605.34  & 153.58     & 83.01     \\
			3 & No action   & 6.38  & sleep  & 54.00 & 60.38  & 497,604.69 & 153.58     & 83.01   \\
			\hline
			\multicolumn{9}{|c|}{Without moving ahead checkpoints} \\
			\hline
			1 & No action   & 4.37  & sleep  & 56.00 & 60.37  & 516.084.73 & 153.59     & 85.82     \\
			2 & No action   & 4.37  & sleep  & 56.00 & 60.38  & 516,085.34  & 153.59     & 85.82     \\
			3 & No action   & 4.38  & sleep  & 56.00 & 60.38  & 516,084.69 & 153.59     & 85.82    \\
			\hline
		\end{tabular}
	}
	\label{tab:checkAdelantado}
	\end{table}

%23_Con_y_sin_checkAdel

\subsubsection{\textbf{Scenario 7: Matrix Multiplication}}

In this scenario, we analyze an MPI master-worker matrix multiplication application ($C = A \times B$), because it is well-known, computationally intensive, and representative application of scientific computing. The Table~\ref{tab:matmulParametros} shows the configuration of the simulation. The test application does successive multiplication of different $A_i$ submatrices  with $B$ to obtain resultant $C_i$. To achieve this, the master process broaadcast matrix $B$ to the worker processes. After that, the master process distributes parts of $A_i$, the worker processes compute the multiplication, and return the partial result to the master process. When all the worker finish this stage, the application enter the next one, where the master distributes another part of $A_i$ until completing $C$. 

The application was characterized with the pas2p tool  \cite{wong2010pas2p}, considering matrices of 2048$\times$2048 elements (of type double) and 4 processes/nodes. As a result, the communication intervals between processes were obtained.

Three different cases are analyzed: The first case shows a long re-execution time with blocking operations, the second case shows a short re-execution time with blocking operations, and the third case shows a short re-execution time with non-blocking operations (the graphics has different time scales). The Table ~\ref{tab:matmulActions} shows the results obtained for these cases.

\begin{table}[t]
\centering
\caption{Simulation parameters for scenario 7:  Matrix multiplication}
\resizebox{\columnwidth}{!}{%
    \begin{tabular}{ |l|l| }
        \hline
        Communication interval & 6.5 sec. \\
        \hline
        Communication pattern & Process P0 sends and receives messages from processes P1, P2, and P3.   \\
        \hline
        MPI Waits & Active     \\
        \hline
        MPI operations & Blocking (first two cases) and non-blocking  (last case) \\
        \hline		
    \end{tabular}
}
	\label{tab:matmulParametros}
\end{table}

\begin{figure*}[h]
	\centering
	\includegraphics[width=\columnwidth,height=0.29\textheight]{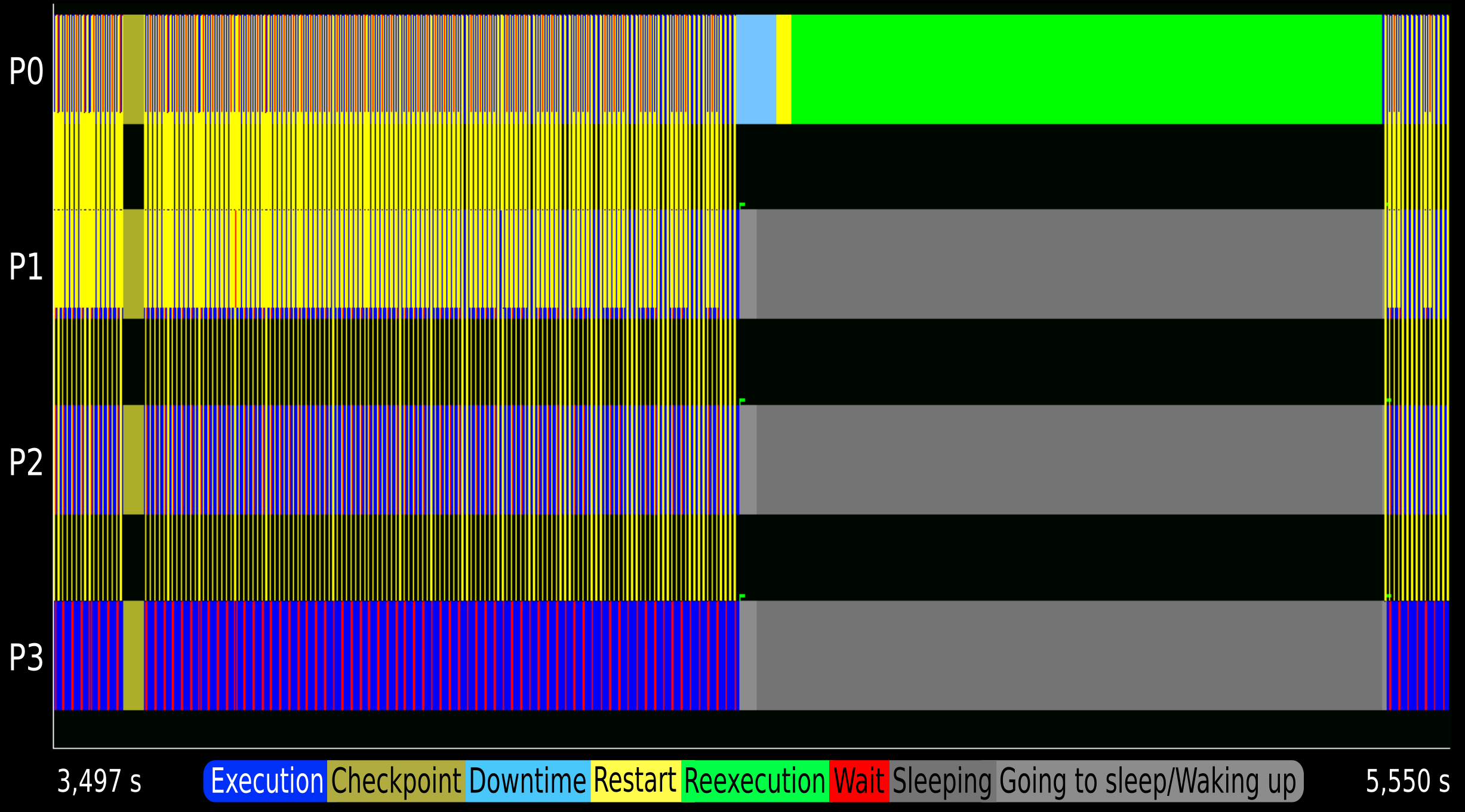}
	\caption{Scenario 7: Matrix multiplication with long reexecution time (blocking operations)}
	\label{fig:matmulMitadIntervalo}
\end{figure*}

In Fig.~\ref{fig:matmulMitadIntervalo}, the case of the long re-execution time from the checkpoint is presented. In this case, the nodes corresponding to the three live processes are sent to sleep, obtaining an estimated saving of 91\% (144.929 J), for an intervention time of approximately 15 minutes. This means that, in the 15 minutes that the failed process needs to recover, the energy consumption would be 91\% lower than the consumption obtained without applying any of the strategies.

%19E_matmul_segundos_falloMitadInter

\begin{figure*}[h]
	\centering
	\includegraphics[width=\columnwidth,height=0.29\textheight]{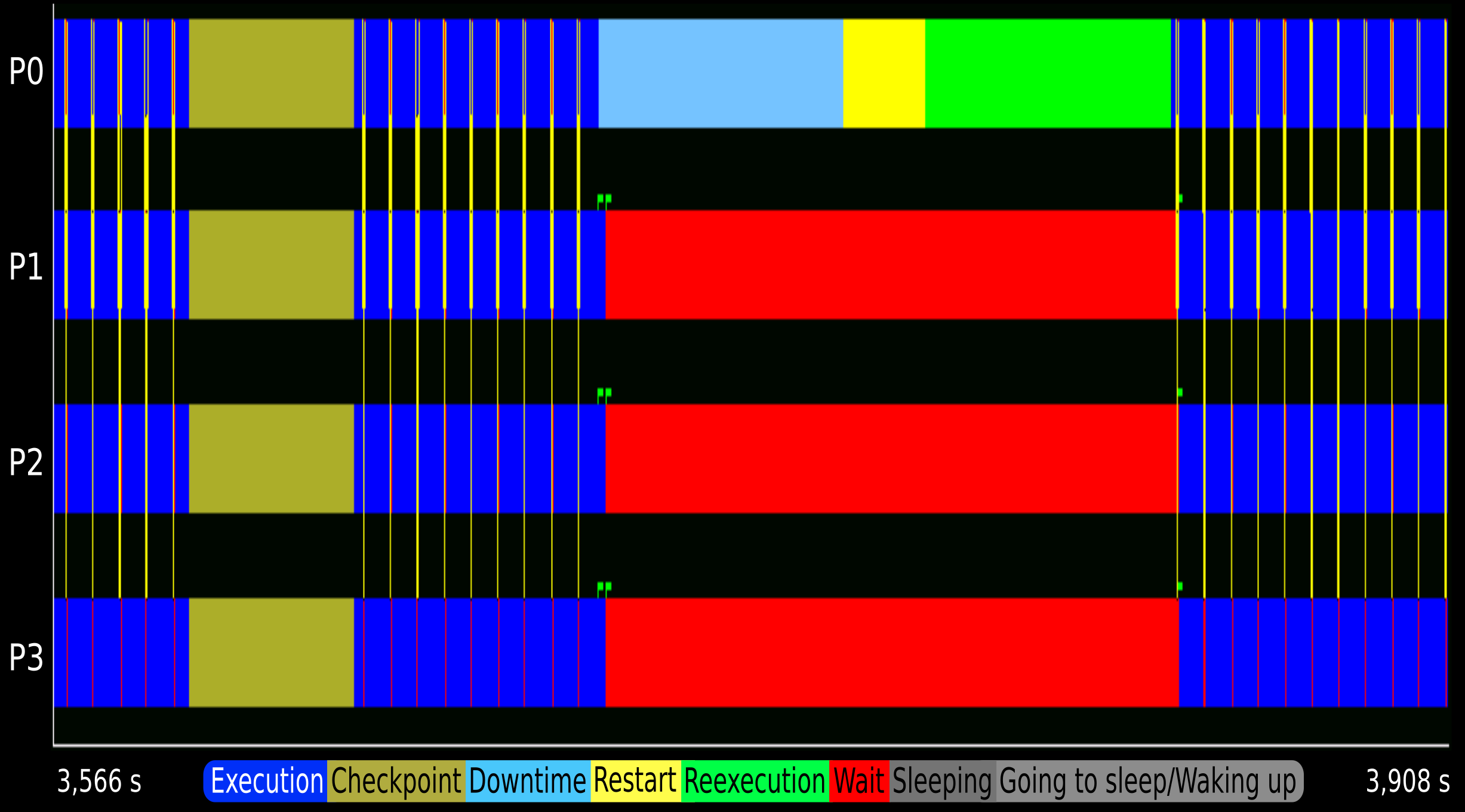}
	\caption{Scenario 7: Matrix multiplication with short reexecution time (blocking operations)}
	\label{fig:matmul}
\end{figure*}

If, on the other hand, the failure occurs near the checkpoint (Fig. \ref{fig:matmul}), the intervention interval is reduced to two and a half minutes, achieving a saving of around 43\% (about 10,000J, against almost 145,000J of the previous case). The selected strategy in this case was to change the frequency of live nodes in its computation and waiting phase.

%21E_matmul_falloCercaCheck

\begin{figure*}[h]
	\centering
	\includegraphics[width=\columnwidth,height=0.29\textheight]{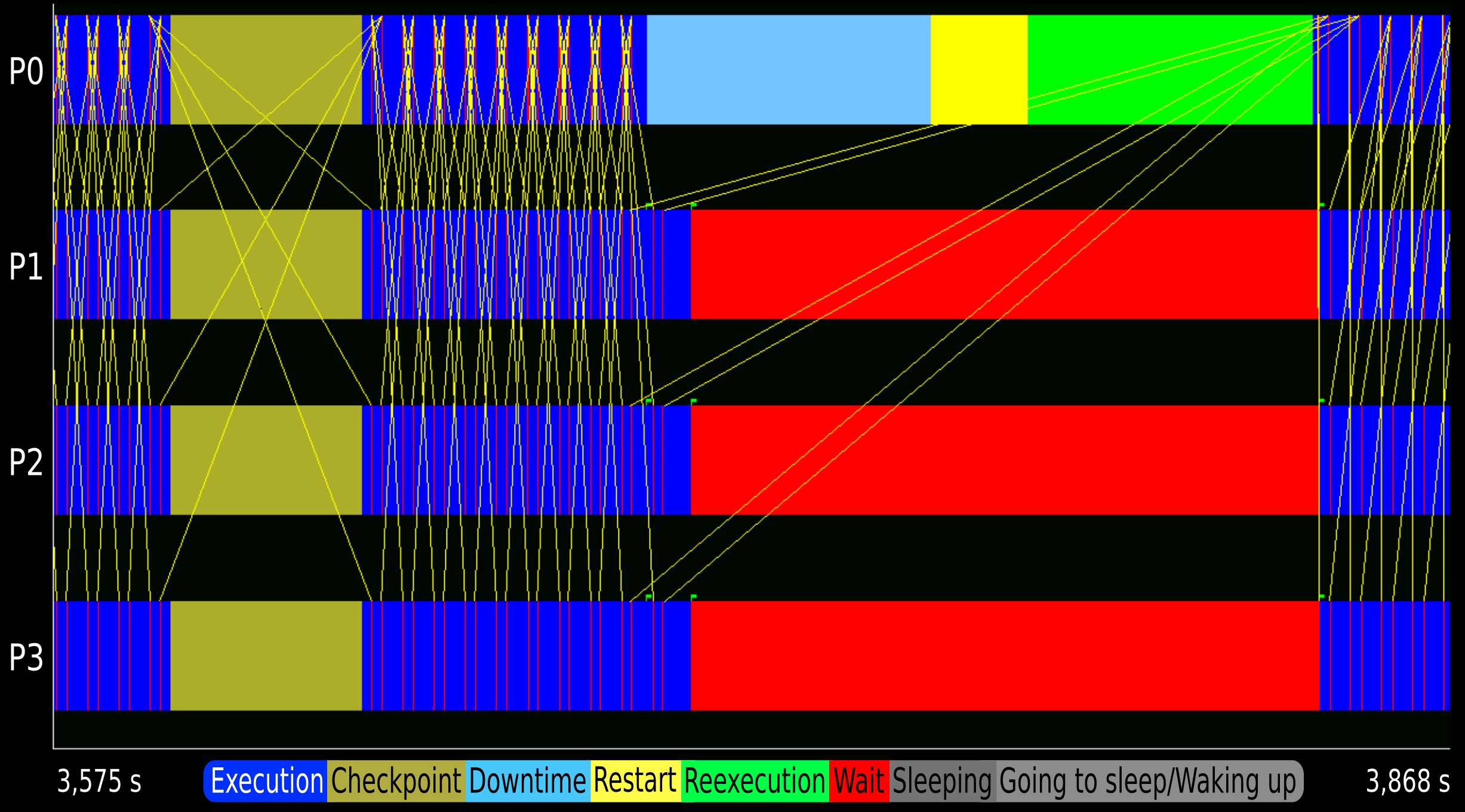}
	\caption{Scenario 7: Matrix multiplication with short reexecution time (non-blocking operations)}
	\label{fig:matmulNoBlock}
\end{figure*}

Fig. \ref{fig:matmulNoBlock} shows the case of short re-execution time with blocking operations. The selected strategy was the same as in the previous case, changing the clock frequency in the computation and waiting phases of all live processes. However, in this case, the computation phase lasts 9 seconds, against the 2 seconds for the same phase in the version with blocking operations. The saving obtained in this case is around 41\% (9,605J, against the 10,022J of the previous case), in the same time interval (two and a half minutes).

%21E_matmul_inmediatas

\begin{table}[t]
  \centering
\caption{Selected actions and energy savings for scenario 7:  Matrix multiplication}
\resizebox{\columnwidth}{!}{%
    \begin{tabular}{ |c|c|r|c|r|r|r|r|r| }
        \hline
        &  \multicolumn{2}{|c|}{Compute phase} &  \multicolumn{2}{|c|}{Wait phase} & \multicolumn{4}{|c|}{}  \\
        \hline
        Node & Action & T (m) & Action & T (m) & TT (m) & Save (J) & Save Rate (J/s) & Save (\%) \\
        \hline
        \multicolumn{9}{|c|}{With long reexecution time (blocking operations)} \\
        \hline
        1 & No action   & 0.09  & sleep  & 15.83 & 15.92  & 144,929.96 & 152.56     & 91.39     \\
        2 & No action   & 0.09  & sleep  & 15.83 & 15.92  & 144,929.94  & 152.56     & 91.39   \\
        3 & No action   & 0.09  & sleep  & 15.83 & 15.92  & 144,929.92 & 152.56     & 91.39  \\
        \hline
        \multicolumn{9}{|c|}{With short reexecution time (blocking operations)} \\
        \hline
        1 & 2.1 GHz   & 0.03  & 1.2 GHz   & 2.33 & 2.36  & 10,022.36 & 70.73     & 42.61    \\
        2 & 2.1 GHz   & 0.03  & 1.2 GHz   & 2.33 & 2.36  & 10,022.40  & 70.73    & 42.61     \\
        3 & 2.1 GHz   & 0.03  & 1.2 GHz   & 2.33 & 2.36  & 10,022.46 & 70.72     & 42.61    \\
        \hline
        \multicolumn{9}{|c|}{With short reexecution time (non-blocking operations)} \\
        \hline
        1 & 2.1 GHz    & 0.16  & 1.2 GHz  & 2.20 & 2.36  & 9,605.29 & 67.92  & 40.92     \\
        2 & 2.1 GHz    & 0.16  & 1.2 GHz  & 2.20 & 2.36  & 9,605.62  & 67.92 & 40.92     \\
        3 & 2.1 GHz    & 0.16  & 1.2 GHz  & 2.20 & 2.36  & 9,605.39 & 67.92  & 40.91    \\
        \hline
    \end{tabular}
}
\label{tab:matmulActions}
\end{table}

\subsection{Discussion}

The experimental work with the simulator shows us that the duration of the waiting phase defines the selection of the strategy. If the wait is long enough to send the node to sleep, this will be the chosen option, since the energy savings obtained exceed by far those obtained with any combination of the other strategies. Furthermore, when the strategy of sleeping a node is selected, the model that the simulator implements always selects, for the compute phase, the lowest clock frequency that does not produce an increase in execution time. In this way, the wait is not shortened, and energy savings are maximized. Moreover, if the failure occurs near the checkpoint time, the wait may be short, and in these cases, the duration of the compute phase plays a fundamental role.

In SPMD or non-dynamic Master-Worker applications where blockings propagate rapidly, it might be counterproductive to change the clock frequency in the computation phase to sleep longer during the waiting phase. In applications with loosely coupled communication patterns, where one group of tasks communicates infrequently with another group of tasks, there can be frequency change during the compute and waiting phases, but this will also depend on whether the re-execution (of the recovering processes) is short.

%% file: secciones/conclusiones.tex
\section{Conclusions and future work}\label{sec:conclusiones}

In this work, we have enriched our previous energy model by including non-blocking communications (with and without system buffering) and the cascade-blocking effect. Next, we have extended the simulator by incorporating the new features of the energy model, to evaluate their potential benefits. 

While implementing the proposal some issues were identified as being important:

 \begin{itemize}
	\item The \textit{depth} of communications, that is, the number of successful communications that a process can perform before it gets blocked by another living process, due to collateral effects of the failure.
	
	\item The slowdown propagation when the clock frequency is changed in the compute phase (in a context with cascade-blocked processes included in the analysis).
	
	\item The need to wake up a node early, because a process on another node needs it awake to continue its execution.
\end{itemize}

Through the experimental work, we were able to discover some opportunities to reduce energy consumption in these scenarios.  The simulations show that the savings were negligible in the worst case, but in some scenarios, it was possible to achieve significant ones; the maximum saving achieved was 90\% in an execution time of 16 minutes.  More generally, we can conclude that:

 \begin{itemize}
 \item Non-blocking operations and the use of system buffering in communications present fewer opportunities for the application of strategies because they avoid blocking states and permit the computation to proceed.
 
 \item Incorporating cascade-blocked processes shows significant increases in energy savings even when just a few nodes are considered.

 \item In applications where communication is unusual, the strategies for the compute phase play a key role (since these phases are long).

\item In applications where compute phases are short, the duration of the waiting phase defines the selection of the strategy. In particular, if the failure occurs far from the last checkpoint time, the waiting phases will be long and the node should probably be sent to sleep.

\end{itemize}

As a result, we show the feasibility of improving energy efficiency in HPC systems in the presence of a failure. Among future works, we plan: 

\begin{itemize}
	\item To implement the strategies on a real system as a proof of concept, considering the secondary effects described before 

	\item To include runtime characterization of the application.
\end{itemize}